\documentclass[aps,prl,preprintnumbers,showpacs,nofootinbib]{revtex4}

\usepackage{graphicx}
\usepackage{dcolumn}
\usepackage{bm}
\begin{document}
\newcommand{\newc}{\newcommand}
\newc{\ra}{\rightarrow}
\newc{\lra}{\leftrightarrow}
\newc{\beq}{\begin{equation}}
\newc{\eeq}{\end{equation}}
\newc{\barr}{\begin{eqnarray}}
\newc{\earr}{\end{eqnarray}}
\newcommand{\Od}{{\cal O}}
\newcommand{\lsim}   {\mathrel{\mathop{\kern 0pt \rlap
  {\raise.2ex\hbox{$<$}}}
  \lower.9ex\hbox{\kern-.190em $\sim$}}}
\newcommand{\gsim}   {\mathrel{\mathop{\kern 0pt \rlap
  {\raise.2ex\hbox{$>$}}}
  \lower.9ex\hbox{\kern-.190em $\sim$}}}

\title{DIRECT SUSY DARK MATTER DETECTION-\\
THEORETICAL RATES DUE TO THE SPIN
 }

\author{J. D. Vergados$^{(1),(2)}$\thanks{Vergados@cc.uoi.gr}}
\affiliation{$^{(1)}${\it Theoretical Physics Division, University
of Ioannina, Ioannina, Gr 451 10, Greece.}}
\affiliation{$^{(2)}${\it T-6, Theoretical Physics Division, LANL,
Los Alamos, N.M. 87545,USA .}}
\begin{abstract}
The recent WMAP data have confirmed that exotic dark matter
together with the vacuum energy (cosmological constant) dominate
in the flat Universe. Thus the direct dark matter detection,
consisting of detecting the recoiling nucleus, is central to
particle physics and cosmology. Supersymmetry provides a natural
dark matter candidate, the lightest supersymmetric particle (LSP).
The relevant cross sections arise out of two mechanisms: i) The
coherent mode, due to the scalar interaction and ii) The spin
contribution arising from the axial current. In this paper we will
focus on the spin contribution, which is expected to dominate for
light targets.
 For both modes it is possible to obtain detectable rates, but in most models  the
 expected rates are much lower than the present experimental goals.
So one should exploit
 two characteristic signatures of the reaction, namely
the modulation effect and, in directional experiments, the
correlation of the event rates with the sun's motion.
 In standard non directional experiments the modulation is small,
less than two per cent. In the case of the directional event rates
we like to suggest that the experiments exploit two features, of
the process, which are essentially independent of the SUSY model
employed, namely: 1) The forward-backward asymmetry, with respect
to the sun's direction of motion, is very large and 2) The
modulation is much larger, especially if the observation is made
in a plane
 perpendicular to the sun's velocity. In this case the difference  between maximum
  and minimum can be  larger than 40 per cent and
 the phase of the Earth at the maximum is direction dependent.
\end{abstract}

\date{\today}
\maketitle

\section{Introduction}
The combined MAXIMA-1 \cite{MAXIMA-1}, BOOMERANG \cite{BOOMERANG},
DASI \cite{DASI}, COBE/DMR Cosmic Microwave Background (CMB)
observations \cite{COBE}, the recent WMAP data \cite{SPERGEL} and
SDSS
 \cite{SDSS} imply that the
Universe is flat \cite{flat01} and
 and that most of the matter in
the Universe is dark, i.e. exotic.
  $$ \Omega_b=0.044\pm 0.04,
\Omega_m=0.27\pm 0.04,  \Omega_{\Lambda}=0.69\pm0.08$$
 for baryonic matter , cold dark matter and dark energy
respectively. An analysis of a combination of SDSS and WMAP data
yields \cite{SDSS} $\Omega_m\approx0.30\pm0.04(1\sigma)$. Crudely
speaking and easy to remember
$$\Omega_b\approx 0.05, \Omega _{CDM}\approx 0.30, \Omega_{\Lambda}\approx 0.65$$

Since the non exotic component cannot exceed $40\%$ of the CDM
~\cite {Benne}, there is room for exotic WIMP's (Weakly
Interacting Massive Particles).
  In fact the DAMA experiment ~\cite {BERNA2} has claimed the observation of one signal in direct
detection of a WIMP, which with better statistics has subsequently
been interpreted as a modulation signal \cite{BERNA1}.

 Supersymmetry naturally provides candidates for the dark matter constituents
\cite{Jung},\cite{GOODWIT}-\cite{ELLROSZ}.
 In the most favored scenario of supersymmetry the
LSP can be simply described as a Majorana fermion, a linear
combination of the neutral components of the gauginos and
higgsinos \cite{Jung},\cite{GOODWIT}-\cite{ref2}. In most
calculations the neutralino is assumed to be primarily a gaugino,
usually a bino. Models which predict a substantial fraction of
higgsino lead to a relatively large spin induced cross section due
to the Z-exchange. Such models have been less popular, since they
tend to violate the relic abundance constraint.
 The upper bound on this constraint has, however, been decreased by the recent
WMAP data. In fact  the LSP relic abundance (including
co-annihilation) is:
\begin{itemize}
\item Before WMAP:
  $$0.09 \leq\Omega_{LSP}h^2\leq0.22$$
\item After WMAP:
$$0.09 \leq\Omega_{LSP}h^2\leq0.124$$
\end{itemize}
These  fairly stringent constrains, however, apply only in the
thermal production mechanism. Furthermore they do not affect the
LSP density in our vicinity derived from the rotational curves.
 We thus feel free to explore the consequences of two recent models \cite{CHATTO},
\cite{WELLS}, which are non-universal gaugino mass models and give
rise to large higgsino components.

\section{The Essential Theoretical Ingredients  of Direct Detection.}
 Even though there exists firm indirect evidence for a halo of dark matter
 in galaxies from the
 observed rotational curves, it is essential to directly
detect \cite{Jung},\cite{GOODWIT}-\cite{KVprd}
 such matter. Such a direct detection, among other things, may also
 unravel the nature of the constituents of dark matter. The
 possibility of such detection, however, depends on the nature of its
 constituents. Here we will assume that such a constituent is the
 lightest supersymmetric particle or LSP.
  Since this particle is expected to be very massive, $m_{\chi} \geq 30 GeV$, and
extremely non relativistic with average kinetic energy $T \approx
50KeV (m_{\chi}/ 100 GeV)$, it can be directly detected
~\cite{Jung}-\cite{KVprd} mainly via the recoiling of a nucleus
(A,Z) in elastic scattering. The event rate for such a process can
be computed from the following ingredients:
\begin{enumerate}
\item An effective Lagrangian at the elementary particle (quark)
level obtained in the framework of supersymmetry as described ,
e.g., in Refs~\cite{ref2,JDV96}. \item A well defined procedure
for transforming the amplitude obtained using the previous
effective Lagrangian from the quark to the nucleon level, i.e. a
quark model for the nucleon. This step is not trivial, since the
obtained results depend crucially on the content of the nucleon in
quarks other than u and d. This is particularly true for the
scalar couplings, which are proportional to the quark
masses~\cite{drees}$-$\cite{Chen} as well as the isoscalar axial
coupling. \item Nuclear matrix elements
\cite{Ress}$-$\cite{DIVA00} obtained with as reliable as possible
many body nuclear wave functions. Fortunately in the most studied
case of the scalar coupling the situation is quite simple, since
then one needs only the nuclear form factor. Some progress has
also been made in obtaining reliable static spin matrix elements
and spin response functions \cite{DIVA00}.
\end{enumerate}
Since the obtained rates are very low, one would like to be able
to exploit the modulation of the event rates due to the earth's
revolution around the sun \cite{DFS86,FFG88}
\cite{Verg98}$-$\cite{Verg01}. In order to accomplish this one
adopts a folding procedure, i.e one has to assume some velocity
distribution
\cite{DFS86,COLLAR92},~\cite{Verg99,Verg01},\cite{UK01}-\cite{GREEN02}
for the LSP. In addition one would like to exploit the signatures
expected to show up in directional experiments, by observing the
nucleus in a certain direction. Since the sun is moving with
relatively high velocity with respect to the center of the galaxy,
one expects strong correlation of such observations with the
motion of the sun \cite {ref1,UKDMC}. On top of this one expects
to see a more interesting pattern of modulation as well.

 The calculation of this cross section  has become pretty standard.
 One starts with
representative input in the restricted SUSY parameter space as
described in the literature for the scalar interaction~\cite{Gomez,ref2}
 (see also Arnowitt
and Dutta \cite{ARNDU}).
 We will only outline here some features entering the
spin contribution. The spin contribution comes mainly via the
Z-exchange diagram, in which case the amplitude is proportional to
$Z^2_3-Z^2_4$ ($Z_3,Z_4$ are the Higgsino components in the
neutralino). Thus in order to get a substantial contribution the
two higgsino components should be large and different from each
other. Normally the allowed parameter space is constrained so that
the neutralino (LSP) is  primarily gaugino, to allow neutralino
relic abundance in the allowed WMAP region mentioned above. Thus
one cannot take advantage of the small Z mass to obtain large
rates. Models with higgsino-like LSP are possible, but then, as we
have mentioned, the LSP annihilation cross section gets enhanced
and the relic abundance $\Omega_{\chi}~h^2$ gets below the allowed
limit.
 It has recently been shown, however, that in the hyperbolic branch
 of the allowed parameter space \cite{CCN03}, \cite{CHATTO} even
 with a higgsino like neutralino the WMAP relic abundance constraint can be
 respected. So, even though the issue may not be satisfactorily settled,
  we feel that it is worth exploiting the spin cross section in the
 direct neutralino detection, since, among other things, it may populate
 excited  nuclear states, if they happen to be so low in energy that they become
 accessible to  the low energy neutralinos. In order to get simple
 estimates of the spin induced neutralino- nucleon cross-section under favorable
 circumstances, to be used as a guide to other slower processes (the directional rates,
 which is the main
 purpose of the present paper, and  transition rates to the excited states to
 be studied elsewhere)
 we will utilize two recently proposed models:
\begin{enumerate}
\item Non-Universal Gaugino mass models  \cite{CHATTO}.
 The gauginos belong to the adjoint of $SU(5)$
, while the Chiral superfields $\Phi$ are in the n-dimensional\\
$(24\times24)_{sym}=\Sigma ~n,~~~~n=1,24,75,200$.\\
 The three
gaugino masses at GUT are expressed in terms a single SUSY
breaking parameter $M_{1/2}$, i.e. $M^{G,n}_i=C^n_i~M_{1/2}$
 The coefficients $C^n_i$ at the GUT scale are given in the table
 \ref{table.Roy}. One, of course, must take into account the renormalization
effects. These are also given in the same table.
The parameter $\mu$ is constrained by:\\
$\mu^2+0.5M_Z^2=r^{(n)}M^2_{1/2}$ with
$r^{(n)}=(2.1,0.3,1.4)\Leftrightarrow n=(1,75,100)$
 \item Anomaly mediated SUSY breaking (AMSB) \cite{WELLS}, inspired, e.g by
superstring  models \cite{BRIGNOLE}. Three simple cases will be
considered:
$$ABSM~(Bino): M_1=\frac{1}{3}\mu~~,~~M_2=\frac{2}{3}\mu$$
$$ABSM~(Wino): M_1=\frac{3}{2}\mu~~,~~M_2=\frac{1}{2}\mu$$
$$ABSM~(Higgsino): M_1=\frac{3}{2}\mu~~,~~M_2=3\mu$$
with $\mu$ constrained by phenomenology alone. The linear relation
between the Higgsino and the gaugino mass may not be a good
approximation. furthermore it is not so easy to satisfy the LSP
relic abundance constraint. We included, however, such a model in
our calculation to compare its predictions on the neutralino
direct detection rate with other models.
\end{enumerate}
 In a given SUSY model one can calculate the amplitudes associated with the
scalar and the spin amplitudes at the quark level.
Going from the quark to the nucleon level, however, is not trivial.
One distinguishes two cases:
\begin{itemize}
\item  The isoscalar axial current\\
 Most of the proton spin is not due to the quark spins
 (proton spin crisis-EMC effect). In fact one finds that the isoscalar and
isovector axial current couplings transform as follows:
 $$f^0_{A}(q) \rightarrow f^0_{A}= g^0_A~f^0_A(q)~~,
 f^1_{A}(q) \rightarrow f^1_{A}= g^0_A~f^1_A(q)~~,
g^0_A\approx0.1~~,~~g^1_A=1.23$$ Where the label (q) defines the
quantity at the quark level. The axial current components
$f_A^0,~f_A^1$ are defined in the standard weak interaction
formalism, see e.g. \cite{JDV96}. In other words they have been
normalized so that
 $\sigma^{spin}_{p,\chi^0}=3(f_A^0+F_A^1)^2 \sigma_0$,
  $\sigma^{spin}_{n,\chi^0}=3(f_A^0-F_A^1)^2 \sigma_0$, $\sigma_0=\frac{1}{2 \pi}(G_F
  m_p)^2$,
  for the proton and neutron spin cross sections
  respectively.

\item The scalar amplitude
\end{itemize}
The relevant amplitude  at the quark level is proportional to the quark mass.
Thus the naive quark model, in which one considers only u and d quarks
in the nucleon, is not even approximately good. In going to the nucleon level
one must compute the matrix element of the quark number operator multiplied
 by its mass, which we express as a fraction of the proton mass, i.e. :
 \beq
    \Big<N|q \bar{q} m_q |N \Big> = f_q m_N
  \label{fq}
  \eeq
Since we are primarily interested in the spin contribution, we will not
  elaborate here on how one obtains the quantities $f_q$
, but we will refer the reader
to the literature \cite{Chen},\cite{Gomez}, \cite{JELLIS01},
\cite{BOTTINO01}, \cite{GGV03}.

 Once the LSP-nucleon cross section is known,
the LSP-nucleus cross section can be obtained. The differential
cross section with respect to the energy transfer $Q$ for a given
LSP velocity $\upsilon$ can be cast in the form
\begin{equation}
d\sigma (u,\upsilon)= \frac{du}{2 (\mu _r b\upsilon )^2}
[(\bar{\Sigma} _{S}F^2(u)
                       +\bar{\Sigma} _{spin} F_{11}(u)]
\label{2.9}
\end{equation}
where we have used a dimensionless variable $u$, proportional to
$Q$, which is found convenient for handling the nuclear form
factor \cite{KVprd} F(u), namely \beq
u=\frac{Q}{Q_0}~~,~~Q_{0}\approx 40 \times A^{-4/3}~MeV.
 \label{uQ}
 \eeq $\mu_r$
is the reduced LSP-nucleus mass and $b$ is (the harmonic
oscillator) nuclear size parameter. In the above expression we
have neglected the small vector and pseudoscalar terms.

Furthermore
\begin{equation}
\bar{\Sigma} _{S} = \sigma^S_{p,\chi^0} A^2 \frac{\mu^2_r}{\mu^2_r(p)}
\label{2.10}
\end{equation}
$\mu_r(p)\approx m_p$ is the LSP-nucleon reduced mass and
\begin{equation}
\bar{\Sigma} _{spin}  =  (\frac{\mu_r}{\mu_r(p)})^2
                           \sigma^{spin}_{p,\chi^0}~\zeta_{spin},
\zeta_{spin}= \frac{1}{3(1+\frac{f^0_A}{f^1_A})^2}S(u)
\label{2.10a}
\end{equation}
$\sigma^{spin}_{p,\chi^0}$ and $\sigma^{s}_{p,\chi^0}$ are the
proton
 cross-sections associated with the spin and the scalar interactions
respectively and
 \beq
S(u)=[(\frac{f^0_A}{f^1_A} \Omega_0(0))^2
\frac{F_{00}(u)}{F_{11}(u)}
  +  2\frac{f^0_A}{ f^1_A} \Omega_0(0) \Omega_1(0)
\frac{F_{01}(u)}{F_{11}(u)}+  \Omega_1(0))^2  \, ]
\label{S(u)}
\eeq
 The precise definition of the spin response functions $F_{ij}$,
  with $i,j=0,1$ isospin indices, which are essentially the "spin form factors"
  normalized to unity at zero momentum transfer, can be found
 elsewhere \cite{DIVA00}.  As we have already mentioned
  the existing experimental limits imply that
 the scalar LSP-nucleon cross section satisfies:
$\sigma^{s}_{p,\chi^0} \le 10^{-5} pb$. The constraint on the
corresponding spin cross-section is less stringent.

Some  static spin matrix elements \cite{DIVA00}, \cite{Ress}, \cite{KVprd}
for some nuclei of interest are given in
table \ref{table.spin}
\section{Rates}
The differential (non directional) rate with respect to the energy
transfer u can be written as:
\begin{equation}
dR_{undir} = \frac{\rho (0)}{m_{\chi}} \frac{m}{A m_N}
 d\sigma (u,\upsilon) | {\boldmath \upsilon}|
\label{2.18}
\end{equation}
 Where   $\rho (0) = 0.3 GeV/cm^3$ is the LSP density in our vicinity,
 m is the detector mass, $m_{\chi}$ is the LSP mass and
$d\sigma(u,\upsilon )$ was given above.\\
 The corresponding directional differential rate, i.e. when only recoiling nuclei
 with non zero velocity in the direction $\hat{e}$  are observed, is given by :
\begin{eqnarray}
dR_{dir} &=& \frac{\rho (0)}{m_{\chi}} \frac{m}{A m_N}
|\upsilon| \hat{\upsilon}.\hat{e} ~\Theta(\hat{\upsilon}.\hat{e})
 ~\frac{1}{2 \pi}~
d\sigma (u,\upsilon)\\
\nonumber & &\delta(\frac{\sqrt{u}}{\mu_r \upsilon
b\sqrt{2}}-\hat{\upsilon}.\hat{e})
 ~~,~ \Theta (x)= \left \{
\begin{array}{c}1~,x>0\\0~,x<0 \end{array} \right \}
 \label{2.20}
\end{eqnarray}

The LSP is characterized by a velocity distribution. For a given
velocity distribution f(\mbox{\boldmath $\upsilon$}$^{\prime}$),
 with respect to the center of the galaxy,
One can find the  velocity distribution in the lab frame
$f(\mbox{\boldmath $\upsilon$},\mbox{\boldmath $\upsilon$}_E)$
by writing

\hspace{2.0cm}\mbox{\boldmath $\upsilon$}$^{'}$=
          \mbox{\boldmath $\upsilon$}$ \, + \,$ \mbox{\boldmath $\upsilon$}$_E
 \,$ ,
\hspace{2.0cm}\mbox{\boldmath $\upsilon$}$_E$=\mbox{\boldmath $\upsilon$}$_0$+
 \mbox{\boldmath $\upsilon$}$_1$

\mbox{\boldmath $\upsilon$}$_0 \,$  is the sun's velocity (around
the center of the galaxy), which coincides with the parameter of
the Maxwellian distribution, and \mbox{\boldmath $\upsilon$}$_1
\,$ the Earth's velocity
 (around the sun).
 The velocity of the earth is given by
\begin{equation}
\mbox{\boldmath $\upsilon$}_E  = \mbox{\boldmath $\upsilon$}_0 \hat{z} +
                                  \mbox{\boldmath $\upsilon$}_1
(\, sin{\alpha} \, {\bf \hat x}
-cos {\alpha} \, cos{\gamma} \, {\bf \hat y}
+ cos {\alpha} \, sin{\gamma} \, {\bf \hat z} \,)
\label{3.6}
\end{equation}
In the above formula $\hat{z}$ is in the direction of the sun`s motion,
$\hat{x}$ is in the radial direction out of the galaxy,  $\hat{y}$ is
perpendicular in the plane of the galaxy ($\hat{y}=\hat{z} \times \hat{x}$)
and $\gamma \approx \pi /6$ is the inclination of the axis of the ecliptic
 with respect to the plane of the galaxy. $\alpha$ is the phase of the Earth
in its motion around the sun ($\alpha=0$ around June 2nd).

The above expressions for the rates must be folded with the LSP velocity
 distribution. We will distinguish two possibilities:
\begin{enumerate}
 \item The direction of the recoiling nucleus is not observed.\\
 The non-directional differential rate is now given by:
\begin{equation}
\Big<\frac{dR_{undir}}{du}\Big> = \Big<\frac{dR}{du}\Big> =
\frac{\rho (0)}{m_{\chi}} \frac{m}{Am_N} \sqrt{\langle
\upsilon^2\rangle } {\langle \frac{d\Sigma}{du}\rangle }
\label{3.11}
\end{equation}
where
\begin{equation}
\langle \frac{d\Sigma}{du}\rangle =\int
           \frac{   |{\boldmath \upsilon}|}
{\sqrt{ \langle \upsilon^2 \rangle}}
 f(\mbox{\boldmath $\upsilon$},\mbox{\boldmath $\upsilon$}_E)
                       \frac{d\sigma (u,\upsilon )}{du} d^3
 \mbox{\boldmath $\upsilon$}
\label{3.12a}
\end{equation}
\item  The direction $\hat{e}$ of the recoiling nucleus is observed.\\
In this case the directional differential rate is given by:
\begin{eqnarray}
\langle (\frac{d\Sigma}{du})_{dir}\rangle &=&\int \frac{
\mbox{\boldmath $\upsilon$}.\hat{e}~
            \Theta( \mbox{\boldmath $\upsilon$}.\hat{e})}
{\sqrt{ \langle \upsilon^2 \rangle}}
 f(\mbox{\boldmath $\upsilon$},\mbox{\boldmath $\upsilon$}_E)
                       \frac{d\sigma (u,\upsilon )}{du}\\
\nonumber & &\frac{1}{2 \pi}
 \delta(\frac{\sqrt{u}}{\mu_r b \upsilon}-\hat{\upsilon}.\hat{e}) d^3
 \mbox{\boldmath $\upsilon$}
\label{3.12b}
\end{eqnarray}
\end{enumerate}
 The above coordinate system, properly taking into account the motion of
 the sun
and the geometry of the galaxy, is not the most convenient for performing
the needed integrations in the case of the directional expressions.
 For this purpose we go to another coordinate
system in which the polar axis, $\hat{Z}$, is in the direction of observation
(direction of the recoiling nucleus) via the transformation:

$\left ( \begin{array}{c}\hat{X} \\
 \hat{Y} \\
\hat{Z} \end{array} \right )=
\left ( \begin{array}{ccc}
\cos{\Theta}\cos{\Phi}&\cos{\Theta}\sin{\Phi}&-\sin{\Theta}\\
-sin{\Phi}&  \cos{\Phi}& 0\\
\sin{\Theta}\cos{\Phi}&\sin{\Theta}\sin{\Phi}&\cos{\Theta}\\
 \end{array} \right )
\left ( \begin{array}{c}\hat{x} \\
\hat{y} \\
\hat{z}
 \label{transf}
 \end{array} \right )$\\
It is thus straightforward to
go to polar coordinates ($ \theta $ , $ \phi $) of the new system
in velocity space and get:
\begin{equation}
\Big<\frac{dR}{du}\Big>_{dir} =\frac{\rho (0)}{m_{\chi}}
\frac{m}{A m_N} \int^{\upsilon_m}_{a \upsilon_0 \sqrt{u}}
\upsilon^3 d \upsilon \int^1_0 \xi d \xi \int^{2 \pi}_0 d \phi
\frac{\tilde{f}({\lambda,\Theta,\Phi, \xi,\phi,\upsilon ,
\upsilon_E )} } {2 \pi}
 \delta(\frac{\sqrt{u}}{\mu_r b \upsilon}-\xi)
                      \frac{d\sigma (u,\upsilon )}{du}
\label{3.26}
\end{equation}
 with $\xi=cos{\theta}$ and $\lambda$ is the asymmetry parameter, which enters
 when one makes the replacement
  $\upsilon_i ^2\Rightarrow(1+\lambda)\upsilon^2_i,~~i=y,~z$ in
 the Maxwell-Boltzmann distribution (with respect to the galactic frame),
  while the x-component remains
 unchanged, see e.g. \cite{ Verg00}.  The orientation parameters $\Theta$
and $\Phi$ appear now explicitly in the distribution function
$\tilde{f}$ and not implicitly via the limits of integration. The
function $\tilde{f}$ can be obtained from the velocity
distribution in a straight forward fashion.
 The $\delta$ function ensures
that in the directional case the variables $u,\upsilon$ and $\xi$
obey the required relation.  In our numerical calculation we found
it more convenient to use $y=\upsilon/\upsilon_0$ and to express
$\xi$ in terms of the other two, namely $u$ and $y$. So one is
left with two integrations, over $\phi$ and $y$. This way, to
leading order in $\delta=2\frac{\upsilon_1}{\upsilon_0}=0.27$,
 we find:
  \beq
 \tilde{f}(\lambda,\Theta,\Phi, \xi,\phi,\upsilon , \upsilon_E)
 \delta(\frac{\sqrt{u}}{\mu_r b \upsilon}-\xi)
 \Rightarrow \tilde{g}(\lambda, \Theta,\Phi,\phi,a\sqrt{u},y,\delta)
 \label{ftilde}
 \eeq
 with
\begin{equation}
 \tilde{g}(\lambda, \Theta,\Phi,\phi,z,y,\delta) =
E(\Theta,\Phi,\lambda,\phi,z,y) \big[
1+\delta~\sin{\alpha}~A(\Theta,\Phi,\phi,z,y))+
\delta~\cos{\alpha}~B(\Theta,\Phi,\lambda,\phi,z,y) \big]
\label{new.eq}
\end{equation}
where $z=a\sqrt{u}$
and a is given by:
  $$a=[\mu_r b \upsilon _0 \sqrt 2 ]^{-1}$$
Furthermore
\begin{eqnarray}
B(\Theta,\Phi,\lambda,\phi,z,y)&=&
(1+\lambda ) \big[-\sin{\gamma}-
\nonumber\\
& & \sin{\gamma} \big(z \cos {\Theta }
-{\sqrt{{y^2}-{z^2}}} \cos {\phi } \sin {\Theta }\big)
+\cos{\gamma}
\big(z \sin {\Theta } \sin {\Phi }+
\nonumber\\
& & {\sqrt{{y^2}-{z^2}}} (\cos {\Phi } \sin {\phi }+
\cos {\Theta } \cos {\phi } \sin {\Phi })\big)\big]
\label{B}
\end{eqnarray}
\begin{eqnarray}
A(\Theta,\Phi,\phi,z,y)= -z \cos{\Phi}  \sin {\Theta }-
{\sqrt{{y^2}-{z^2}}} (\cos {\Theta } \cos {\phi }
 \cos {\Phi }-\sin {\phi }
\sin {\Phi })
\label{A}
\end{eqnarray}
\begin{eqnarray}
E(\Theta,\Phi,\phi,z,y)=
e ^{-(1+\lambda ) \bigg(1+{y^2}+
2 \bigg(z \cos {\Theta }-{\sqrt{{y^2}-{z^2}}}
 \cos {\phi } \sin {\Theta }\bigg)\bigg)} \times
\nonumber\\
e ^{ \lambda  {{\bigg(-z \cos {\Phi }
 \sin {\Theta }-{\sqrt{{y^2}-{z^2}}} (\cos {\Theta } \cos
{\phi } \cos {\Phi }-\sin {\phi } \sin {\Phi })\bigg)}^2}}
\label{E}
\end{eqnarray}
\begin{figure}
\begin{center}
\includegraphics[height=.3\textheight]{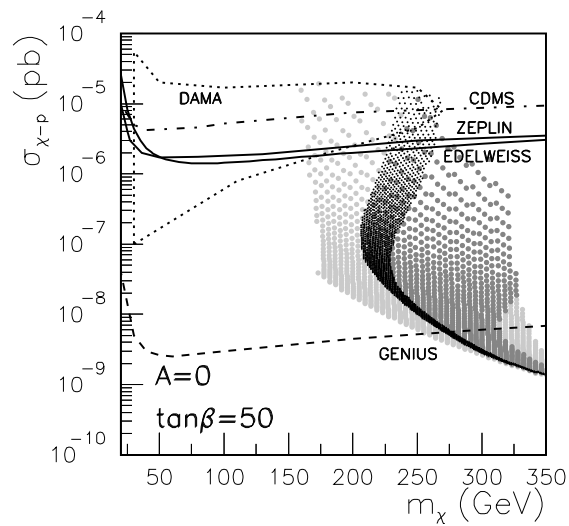}
\caption{
Scatter plots associated with the scalar interaction with non-universal parameters obtained
Cerdeno, Gabrielli, Gomez and Munoz.
The light grey dotted  area corresponds to the  laboratory constraints, while
the dark dotted area is associated with  old relic abundance constraint and
the black  dotted area with the WMAP relic abundance constraint.
\label{mario4} }
\end{center}
\end{figure}
\begin{figure}
\begin{center}
\includegraphics[height=.3\textheight]{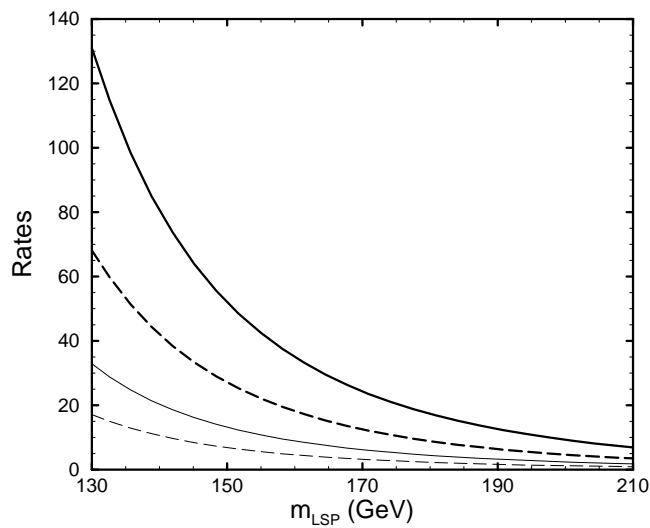}
\caption{The Total detection rate per $(kg-target)yr$ vs the LSP mass
in GeV for  a typical solution in our parameter space in the case of
 $^{127}I$.
Thick lines  correspond to  model B, while fine lines to model C.
 In the upper curve no detector cutoff was employed,
while in the lower curve we used a detector energy cutoff of $Q_{min}=10~KeV$.
Such effects introduce variations in the rates by  factors of about two.
\label{rate}
}
\end{center}
\end{figure}
To obtain the total rates one must integrate  the expressions
\ref{3.12a} and \ref{3.12b}  over the energy transfer from
$Q_{min}$ determined by the detector energy cutoff to $Q_{max}$
determined by the maximum LSP velocity (escape velocity, put in by
hand in the Maxwellian distribution), i.e.
 \beq
\upsilon_{esc}=2.84~\upsilon_0~~, ~~\upsilon_0=229~Km/s.
 \label{vesc}
 \eeq
 In our analysis we included only the rotational velocity of the
 sun around the center of the galaxy. The local component of the
 sun`s velocity is 10 times smaller and it can be neglected.
\section{Results}
We will specialize  the above results in the following cases:
\subsection{Non directional unmodulated rates}
Ignoring the motion of the Earth the total non directional rate is
given by
\begin{equation}
R =  \bar{R}\, t(a,Q_{min}) \,~~,~~\bar{R}=\frac{\rho
(0)}{m_{\chi^0}} \frac{m}{Am_p}~
              (\frac{\mu_r}{\mu_r(p)})^2~ \sqrt{\langle
v^2 \rangle } [\sigma_{p,\chi^0}^{S}~A^2+
 \sigma _{p,\chi^0}^{spin}~\zeta_{spin}]
  \label{3.55f}
\end{equation}
where $t$ is the ratio of the calculated rate divided by that
obtained using the previous equation. It represents the
modification of the total rate due to the
 folding procedure and the nuclear structure effects. $t$  depends on
$Q_{min}$, i.e.  the  energy transfer cutoff imposed by the detector
 and the parameter $a$ introduced above.
 All SUSY parameters, except the LSP mass, have been absorbed in $\bar{R}$.

Via  Eq. (\ref{3.55f}) we can, if we wish,   extract the nucleon cross
 section from the data.
For most of the allowed parameter space the obtained results for
the coherent mode are  undetectable in the current experiments. As
it has already been mentioned it is possible to obtain detectable
rates. Such results have, e.g. recently been obtained by Cerdeno
{\it et al} \cite{CERDENO} with non universal set of parameters
 and the Florida group \cite{BAER03}.

 A representative set is shown in
Fig. \ref{mario4} or
with universal couplings \cite {Gomez} for large $\tan{\beta}$ in
 Fig. \ref{rate} in the case of the target$^{127}I$. The planned experiments,
like CDMS \cite{CDMS}, EDELWEISS \cite{EDELWEISS}, IGEX \cite{IGEX},
ZEPLIN \cite{ZEPLIN} and GENIUS \cite{GENIUS}, are
however,  expected to improve so that they may detect rates two or three
orders of magnitude smaller (see Fig. \ref{mario4}).

\begin{figure}
\begin{center}
\includegraphics[height=.2\textheight]{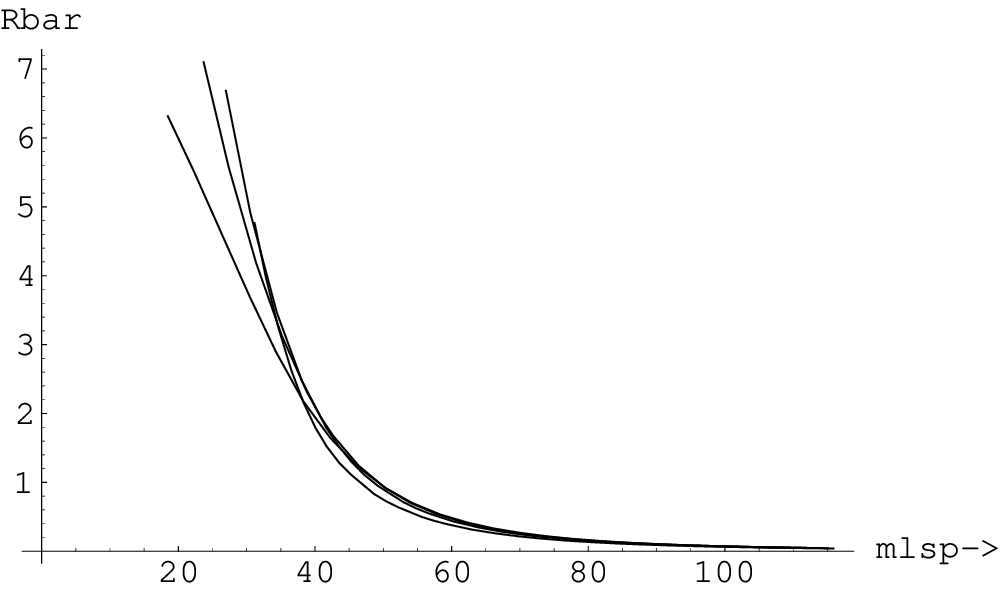}
\includegraphics[height=.2\textheight]{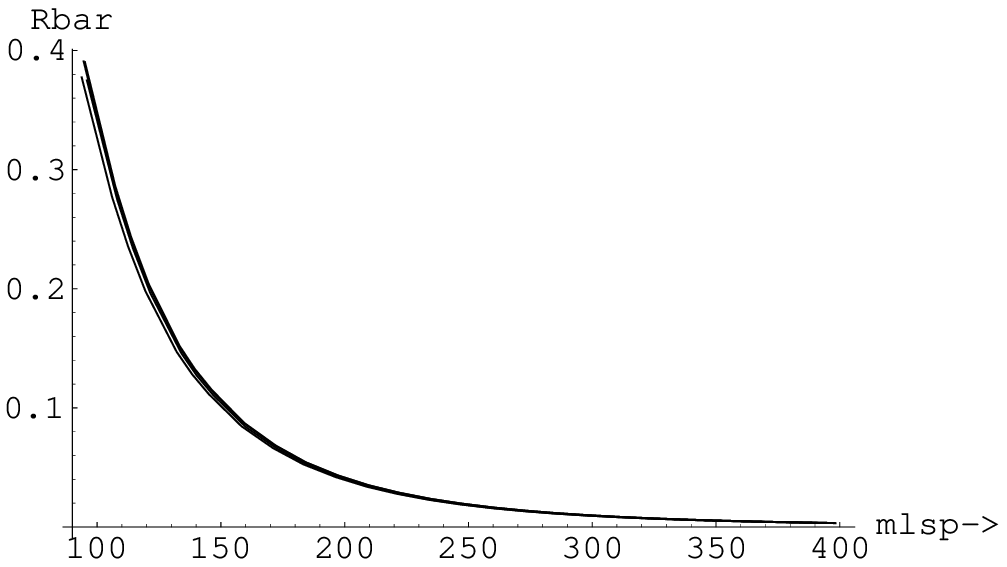}
\caption{
 The quantity $\bar{R},~\approx$ event rate for $Q_{min}=0$,
 associated with  the spin
contribution in the case of the $A=19$ system for model n=1 of
Chattopadhyay and Roy on the left and model n=75 on the right.
  The curves from top to bottom correspond
to ($sign(\mu),tan(\beta)$) :  $(-,40),(+,40),(-,10),(+,10)$. The
curves for $\pm10$ cannot be distinguished from each other.
\label{draw1.19} }
\end{center}
\end{figure}

\begin{figure}
\begin{center}
\includegraphics[height=.2\textheight]{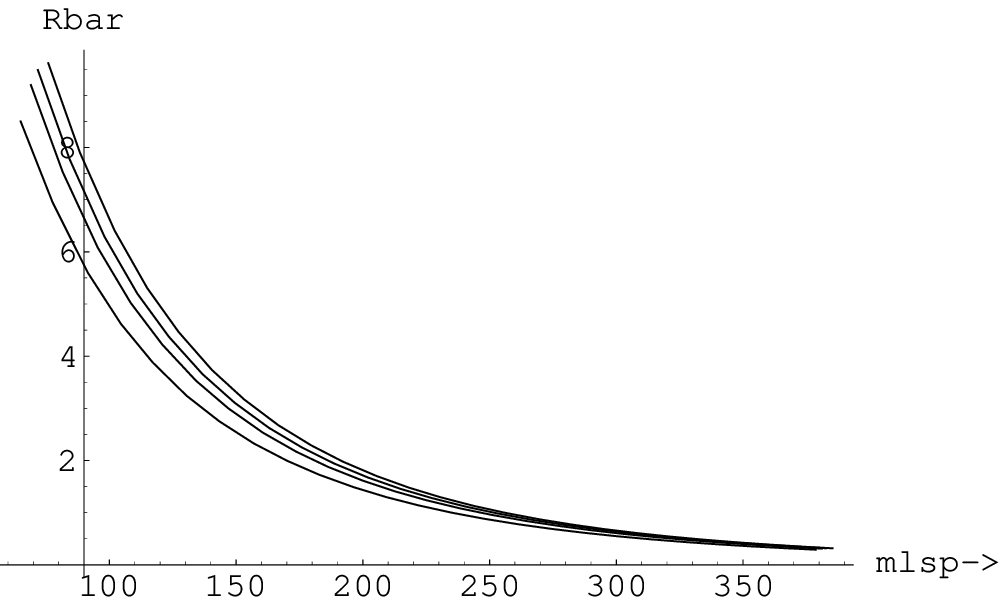}
\includegraphics[height=.2\textheight]{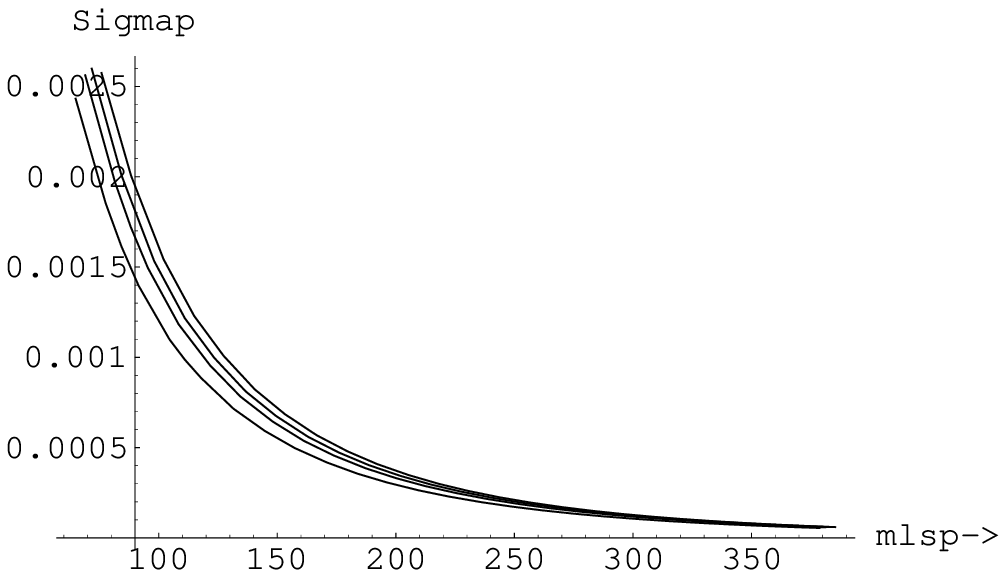}
\caption{ On the left the same as in Fig. \ref{draw1.19} and on
the right the proton cross section in $pb$, both  for $n=100$.
 Since the isoscalar contribution is negligible, the neutron cross section is similar.
 \label{draw2.19} }
\end{center}
\end{figure}

\begin{figure}
\begin{center}
\includegraphics[height=.2\textheight]{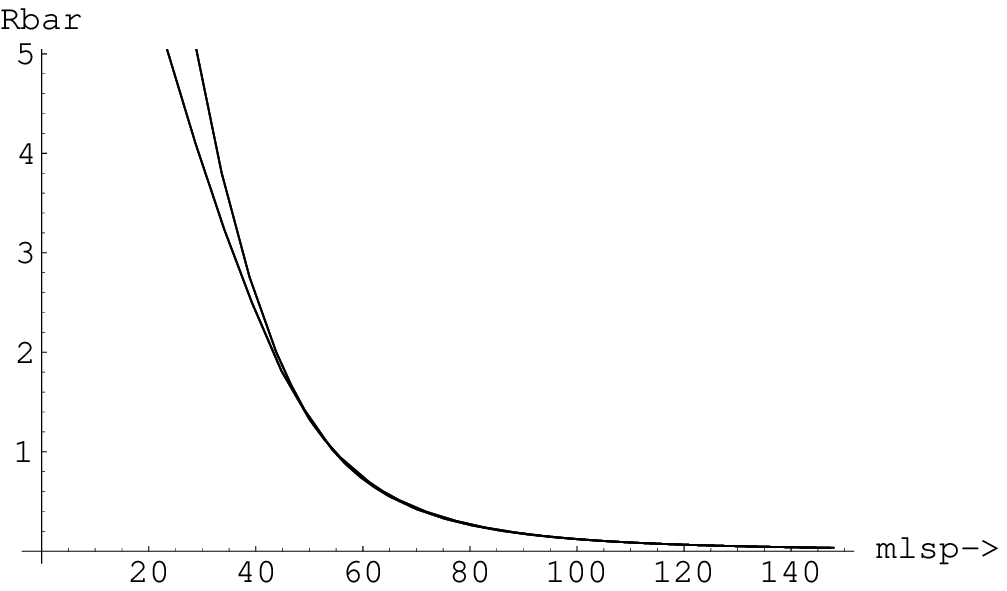}
\includegraphics[height=.2\textheight]{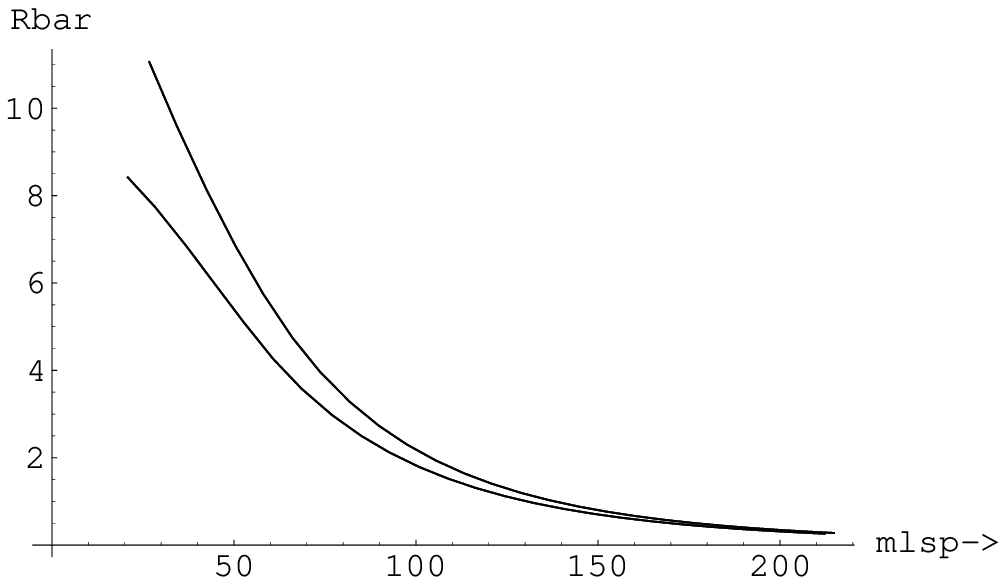}
\caption{
 The quantity $\bar{R},~\approx$ event rate for $Q_{min}=0$, associated
 with  the spin
contribution in the case of the $A=19$ system for model $M_1=(1/3)
\mu,M_2=(2/3)\mu$ of Murakami and Wells
 on the left and model $M_1=(3/2) \mu,M_2=(1/2)\mu$ on the
 right.
  The curves are independent of the sign of $\mu$. From top to bottom correspond
to $tan(\beta): $ $40,10$.
 \label{draw3.19}
 }
\end{center}
\end{figure}

\begin{figure}
\begin{center}
\includegraphics[height=.2\textheight]{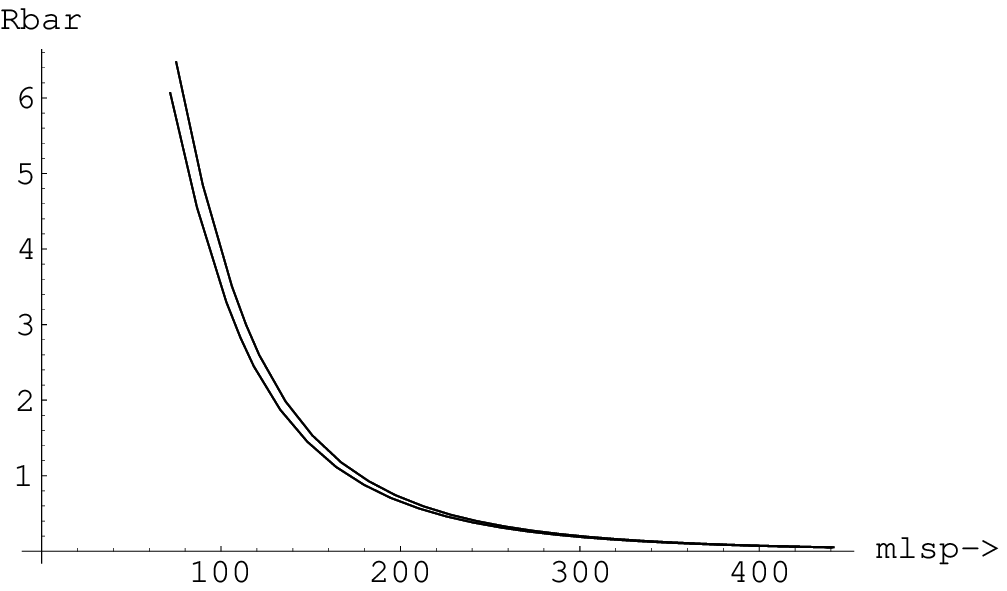}
\caption{The same as in Fig. \ref{draw3.19} for model
$M_1=(3/2) \mu,M_2=3\mu$ of Murakami and Wells. Note the
difference in the scale of the LSP mass compared
to Fig. \ref{draw3.19}.
 \label{draw4.19} }
\end{center}
\end{figure}

In the case of the spin contribution we see that detectable rates
are possible only in those cases in which the neutralino has a
substantial Higgsino component (see Figs \ref{draw1.19}-
\ref{draw2.19} and Figs \ref{draw3.19}- \ref{draw4.19}). Our
results for the nucleon cross section, obtained by neglecting the
isoscalar axial current due to the EMC effect, are in essential
agreement with more detailed calculations, which have appeared
after our manuscript was completed \cite{CCN03}.

\begin{figure}
\begin{center}
\includegraphics[height=.2\textheight]{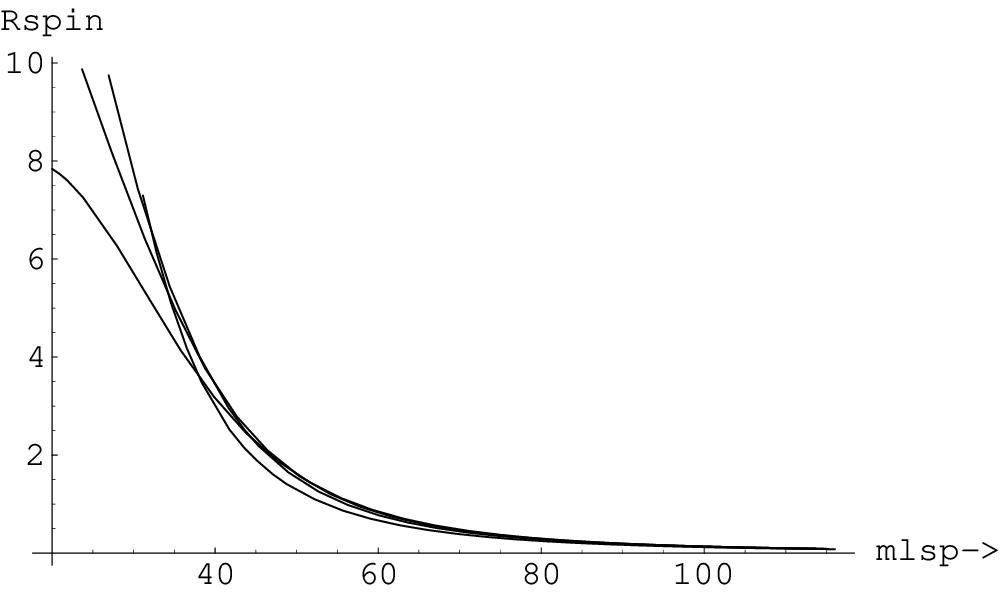}
\includegraphics[height=.2\textheight]{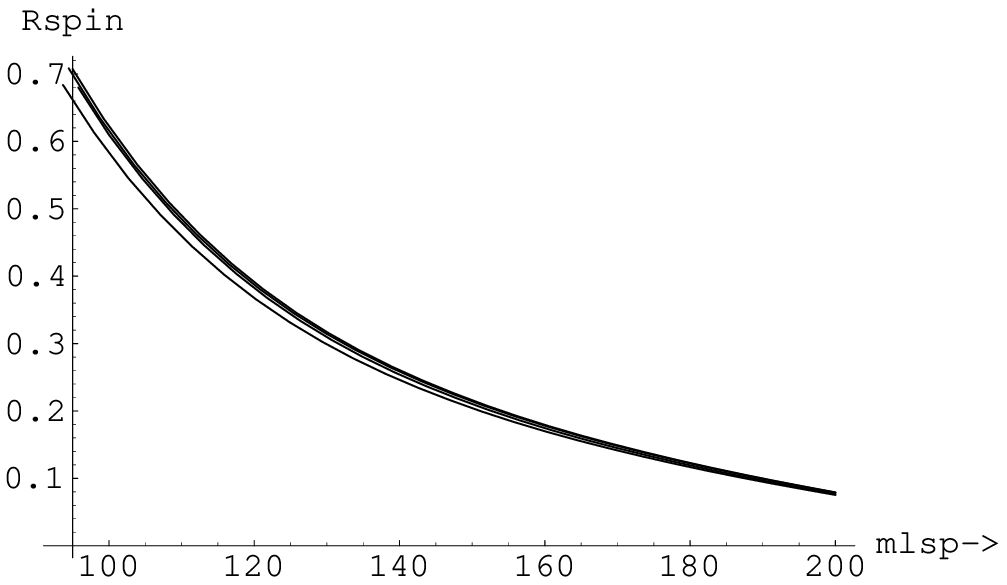}
\caption{
 The event rate, associated with  the spin
contribution in the case of the $A=127$ system. The notation is the same
as in Fig. \ref{draw1.19}.  Only the spherically
symmetric velocity distribution ($\lambda=0$) has been considered.
\label{draw1.127} }
\end{center}
\end{figure}
\begin{figure}
\begin{center}
\includegraphics[height=.2\textheight]{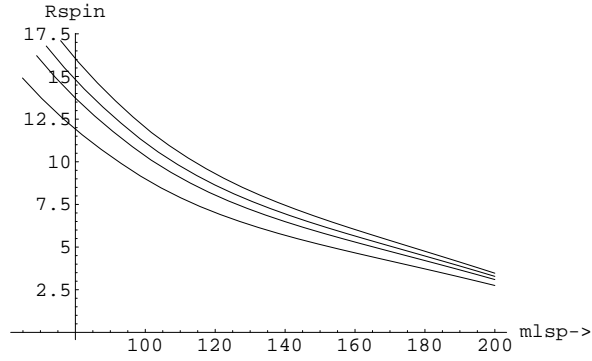}
\caption{ The same as in Fig. \ref{draw1.127} for model n=100.
 \label{draw2.127} }
\end{center}
\end{figure}
\begin{figure}
\begin{center}
\includegraphics[height=.2\textheight]{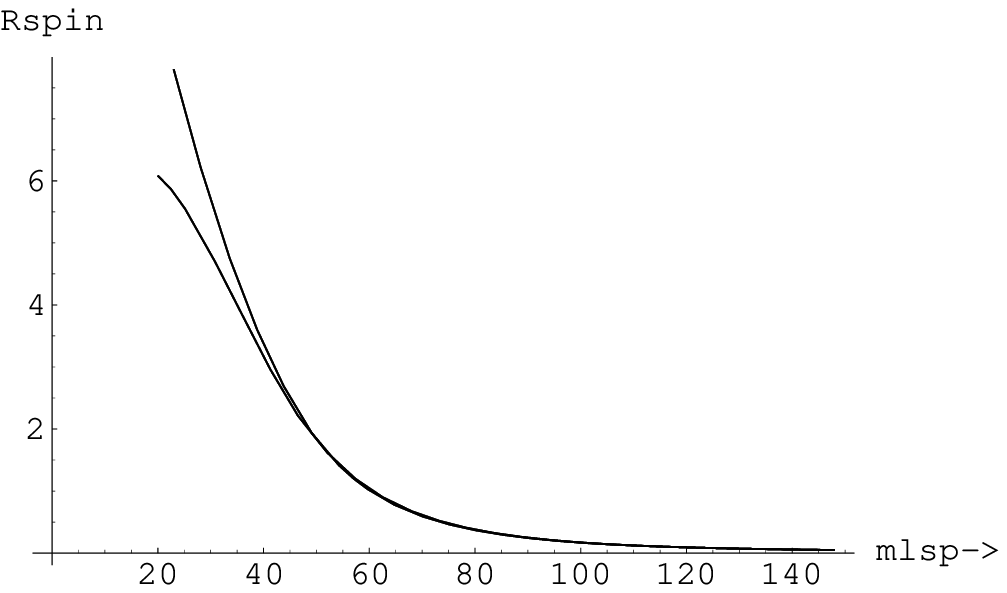}
\includegraphics[height=.2\textheight]{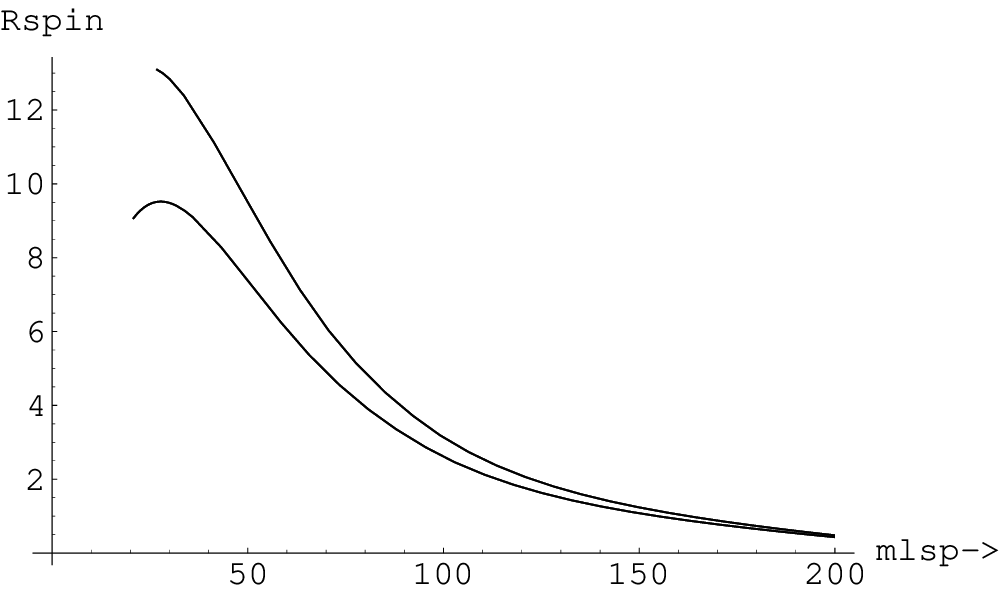}
\caption{
 The event rate, associated with  the spin
contribution in the case of the $A=127$ system for model$M_1=(1/3) \mu,M_2=(2/3)\mu$
of Murakami and Wells
 on the left and model $M_1=(3/2) \mu,M_2=(1/2)\mu$ on the
 right.
  The curves are independent of the sign of $\mu$. From top to bottom correspond
to $tan(\beta): $ $40,10$.
 \label{draw3.127}
 }
\end{center}
\end{figure}
\begin{figure}
\begin{center}
\includegraphics[height=.2\textheight]{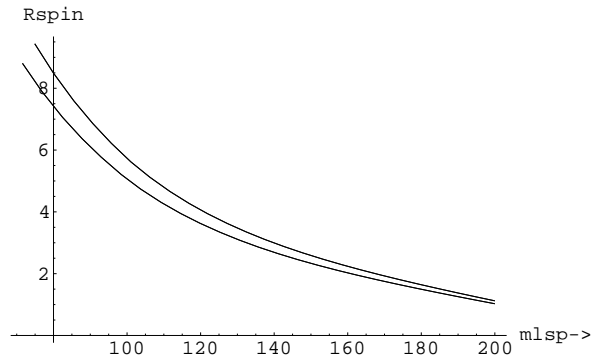}
\caption{The same as in Fig. \ref{draw3.19} for model
$M_1=(3/2) \mu,M_2=3\mu$ of Murakami and Wells. Note the
difference in the scale of the LSP mass compared
to Fig. \ref{draw3.127}.
 \label{draw4.127} }
\end{center}
\end{figure}

For the target $^{127}I$ the corresponding results are shown in
Figs \ref{draw1.127}- \ref{draw2.127} and Figs \ref{draw3.127}-
\ref{draw4.127} in the case of the spherically symmetric M.B.
velocity distribution and $Q_{min}=0$. The effects of
 asymmetry and the detector
energy cutoff $Q_{min}$ are analogous to those previously found in the coherent
mechanism.
\subsection{Modulated Rates}.

If the effects of the motion of the Earth around the sun are included, the total
 non directional rate is given by
\begin{equation}
R =  \bar{R}\, t \, [(1 +  h(a,Q_{min})cos{\alpha})]
\label{3.55j}
\end{equation}
with  $h$  the modulation amplitude, relative to the unmodulated
(time averaged) amplitude, and $\alpha$ is the phase of the Earth,
which is zero around June 2nd. The modulation amplitude would be
an excellent signal in discriminating against background, but
unfortunately it is very small, less
 than two per cent (see table \ref{table1.gaus}).
 Furthermore for intermediate and heavy nuclei, it can even change sign
for sufficiently  heavy LSP. So in our opinion a better signature is provided
 by directional experiments, which measure the direction of the recoiling nucleus.
\subsection{Directional Rates.}
\begin{figure}
\begin{center}
\rotatebox{90}{\hspace{1.0cm} $t_{dir}\rightarrow$}
\includegraphics[height=.15\textheight]{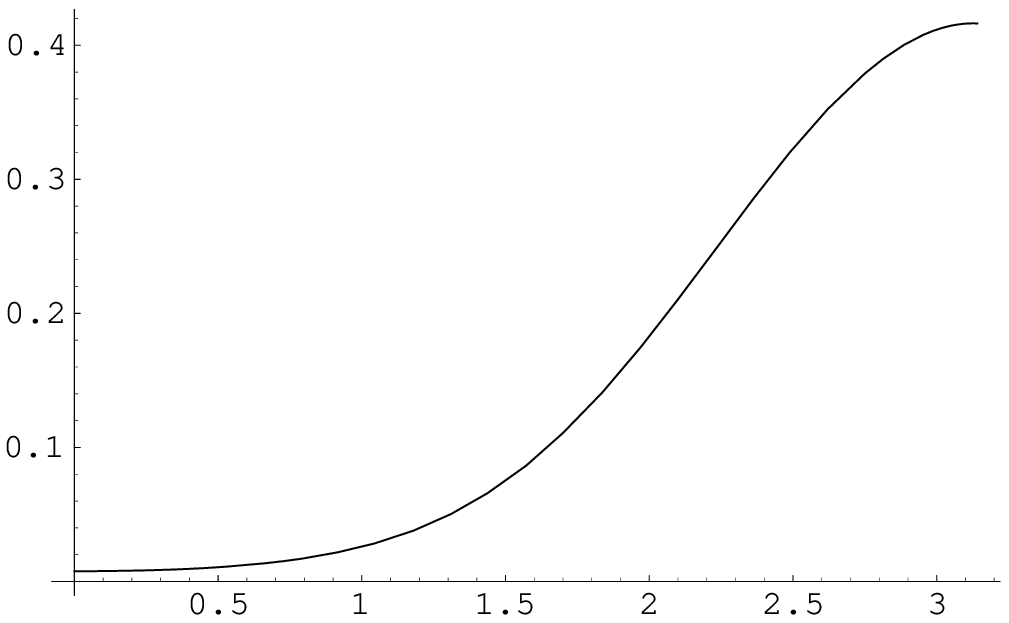}
\hspace{0.0cm} $\Theta \rightarrow$
\rotatebox{90}{\hspace{1.0cm} $t_{dir}\rightarrow$}
\includegraphics[height=.15\textheight]{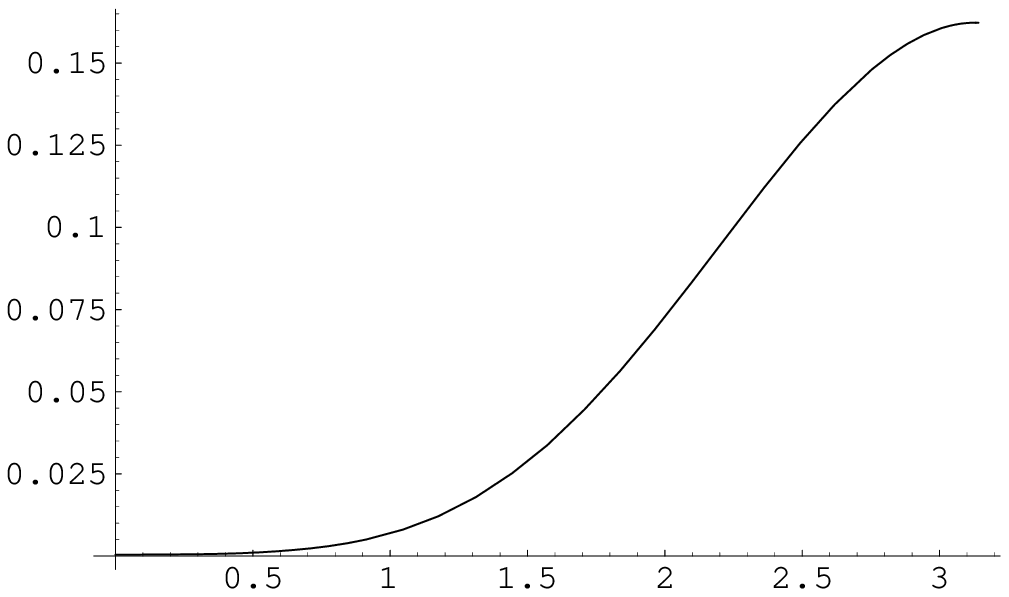}
\hspace{0.0cm} $\Theta \rightarrow$\\
\rotatebox{90}{\hspace{1.0cm} $t_{dir}\rightarrow$}
\includegraphics[height=.15\textheight]{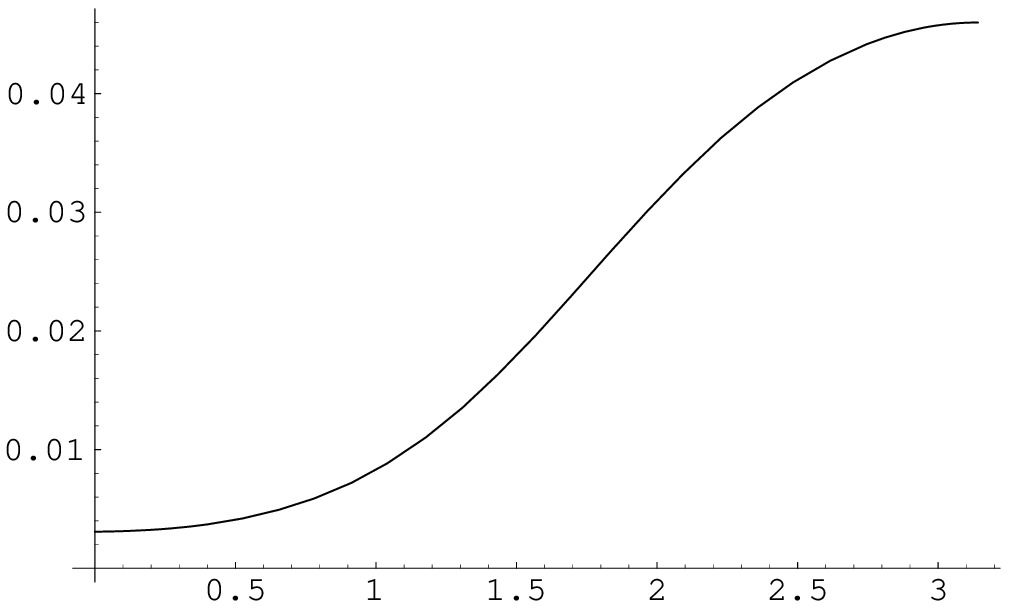}
\hspace{0.0cm} $\Theta \rightarrow$ \rotatebox{90}{\hspace{1.0cm}
$t_{dir}\rightarrow$}
\includegraphics[height=.15\textheight]{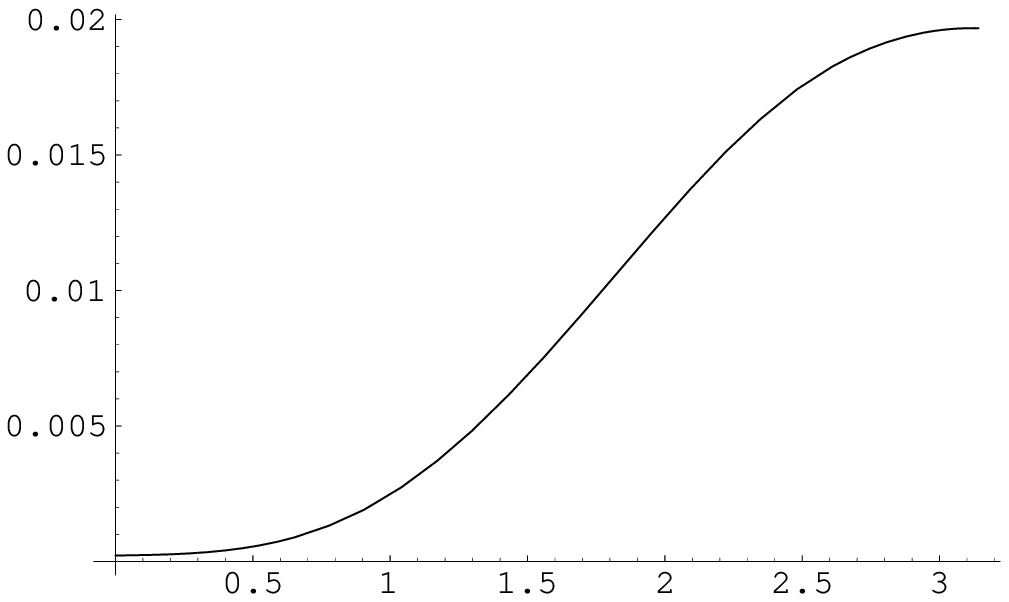}
\hspace{0.0cm} $\Theta \rightarrow$
 \caption{ The parameter
$t_{dir}$ defined in the text as a function of the angle $\Theta$
for $A=19$ at the top and $A=127$ at the bottom. On the left we
show $t_{dir}$ for $\lambda=0$ and on the right for $\lambda=1$.
 The parameter
$t_{dir}$ is independent of $\Phi$ for $\lambda=0$ and it depends slightly
on it, when $\lambda$ is different from zero.
The results presented correspond to an LSP mass of $100~GeV$.
 \label{tdir}
 }
\end{center}
\end{figure}

Since the sun is moving around the galaxy in a directional experiment, i.e. one in which the
direction of the recoiling nucleus is observed, one expects a strong correlation of the
event rate with the motion of the sun. In fact
the directional rate can be written as:
\begin{equation}
R_{dir}  =   \frac{t_{dir}} {2 \pi} \bar{R}  \,
            [1 + h_m  cos {(\alpha-\alpha_m~\pi)}]
          = \frac{\kappa} {2 \pi} \bar{R}~t  \,
            [1 + h_m  cos {(\alpha-\alpha_m~\pi)}]
\label{4.56b}
\end{equation}
where $t_{dir}$ is a quantity analogous to $t$ discussed above,
and $h_m$ is the modulation.
 $\alpha_m $ is the "shift" in the phase of the Earth $\alpha$,
 since now we have both sine and cosine terms and the maximum
 occurs at $\alpha=\alpha_m \pi$. $\kappa/(2 \pi)$ is the reduction
 factor of the unmodulated
directional rate relative to the non-directional one. The
parameters  $\kappa~,~h_m~,~\alpha_m$ depend on the direction of
 observation:
$$\hat{e}=(\sin{\Theta} \cos{\Phi} ~,~\sin{\Theta} \sin{\Phi}~,~\cos{\Theta})$$
 The parameter $t_{dir}$ for a typical LSP mass $100~GeV$ is shown in
 Fig. \ref{tdir} as a function of the angle $\Theta$ for the targets $A=19$
 and $A=127$.
 We see that the change of the rate as a function of
 the angle $\Theta$ for the Maxwellian LSP velocity distribution is quite
 dramatic. This figure is  important in the analysis of the angular
 correlations, since, among other things, there is always un uncertainty in
 the determination of the angle in a directional experiment.
\begin{figure}
\begin{center}
\rotatebox{90}{\hspace{1.0cm} $h_m \rightarrow$}
\includegraphics[height=.15\textheight]{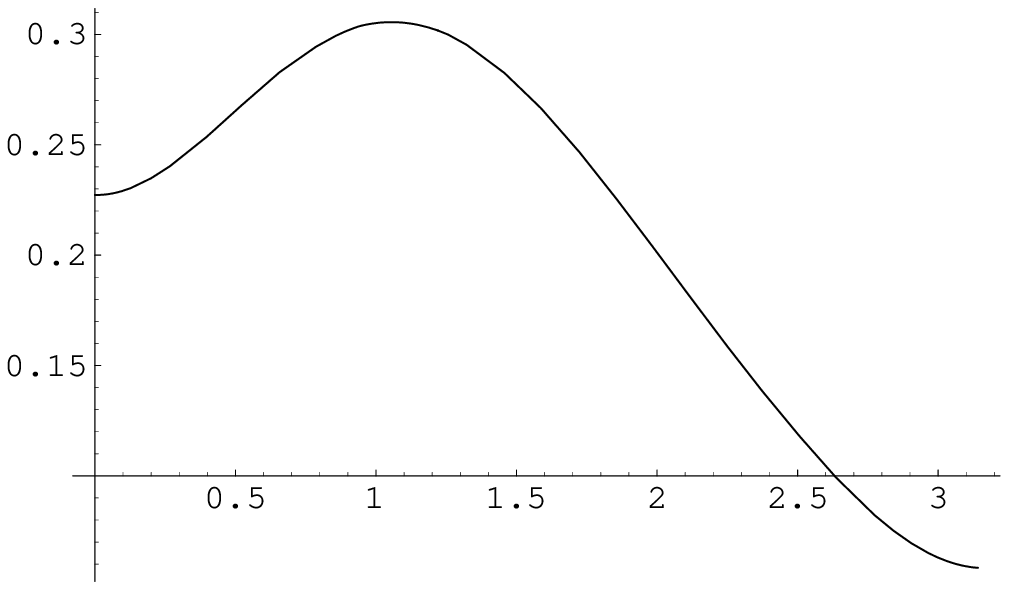}
\hspace{0.0cm} $\Theta \rightarrow$
 \rotatebox{90}{\hspace{1.0cm} $h_m \rightarrow$}
\includegraphics[height=.15\textheight]{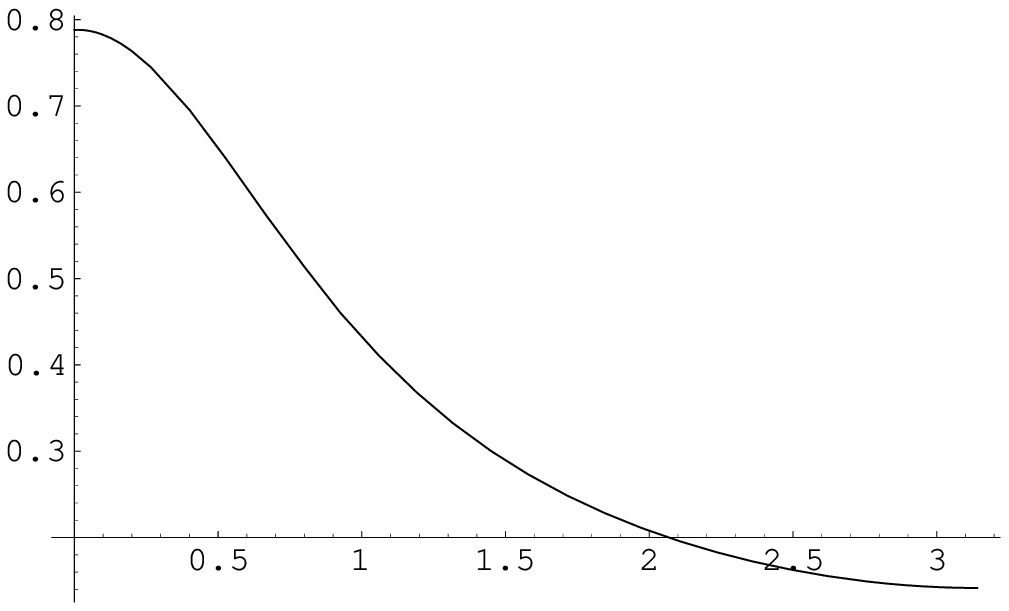}
\hspace{0.0cm} $\Theta \rightarrow$\\
 \rotatebox{90}{\hspace{1.0cm} $h_m \rightarrow$}
\includegraphics[height=.15\textheight]{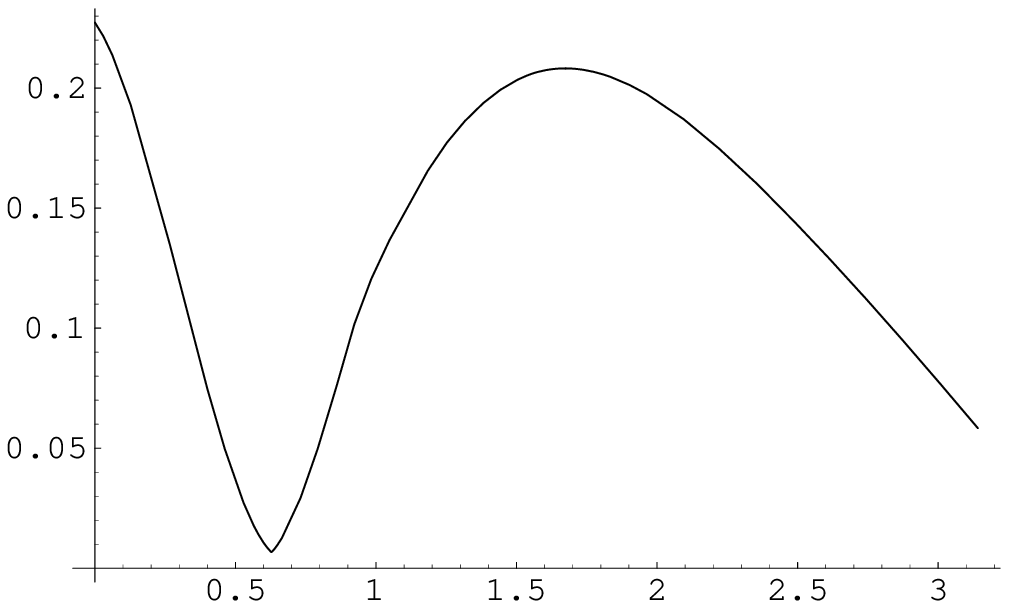}
\hspace{0.0cm} $\Theta \rightarrow$
  \rotatebox{90}{\hspace{1.0cm} $h_m \rightarrow$}
\includegraphics[height=.15\textheight]{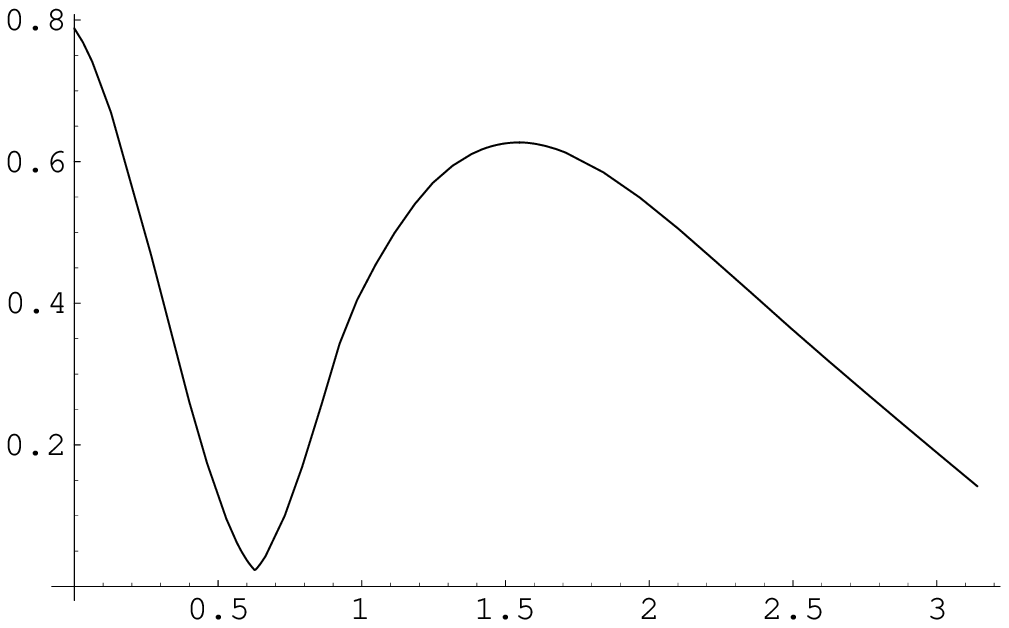}
\hspace{0.0cm} $\Theta \rightarrow$\\
 \rotatebox{90}{\hspace{1.0cm} $h_m \rightarrow$}
\includegraphics[height=.15\textheight]{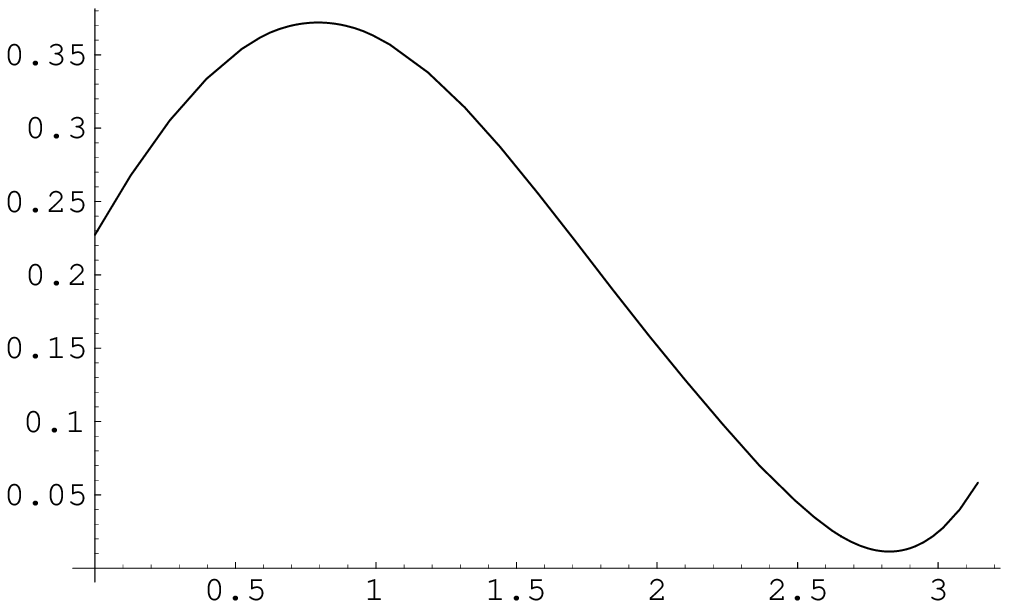}
\hspace{0.0cm} $\Theta \rightarrow$ \rotatebox{90}{\hspace{1.0cm}
$h_m \rightarrow$}
\includegraphics[height=.15\textheight]{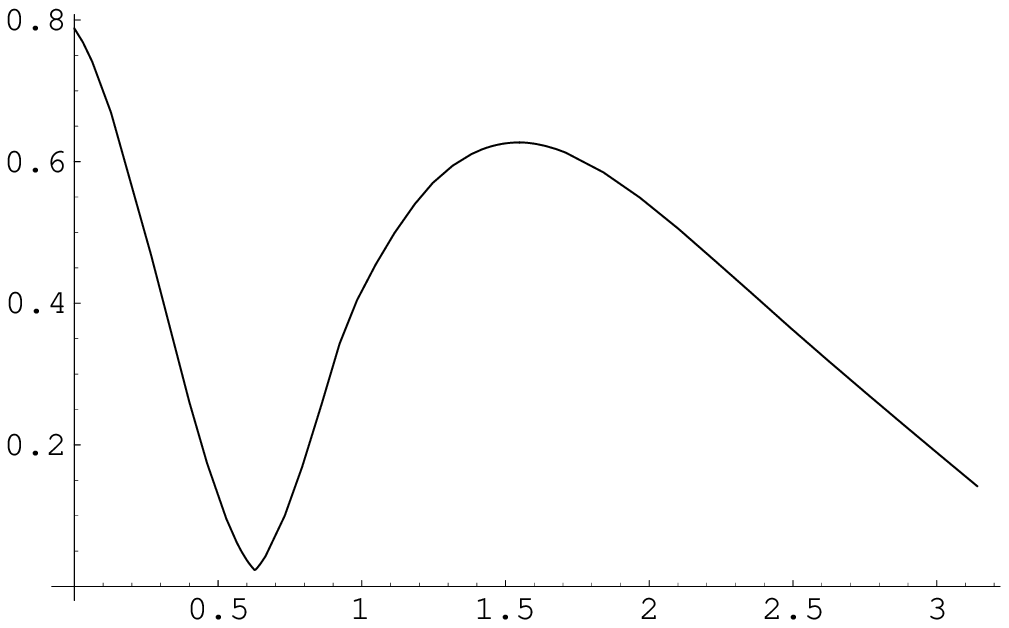}
\hspace{0.0cm} $\Theta \rightarrow$
\caption{ The modulation amplitude $h_m$ in the case of the $A=19$ system as
function of $\Theta$ for $\Phi=0,2\pi$ at the top,
 $\Phi=\pi/2$ in the middle
 and $\Phi=3~\pi/2$ at the bottom. The values for angles $\Theta \le 0.5$
should be discarded since the rate is tiny (see Fig. \ref{tdir}).
Otherwise the notation is the same as in Fig \ref{tdir}.
 \label{hdir19}
 }
\end{center}
\end{figure}
\begin{figure}
\begin{center}
\rotatebox{90}{\hspace{1.0cm} $h_m \rightarrow$}
\includegraphics[height=.15\textheight]{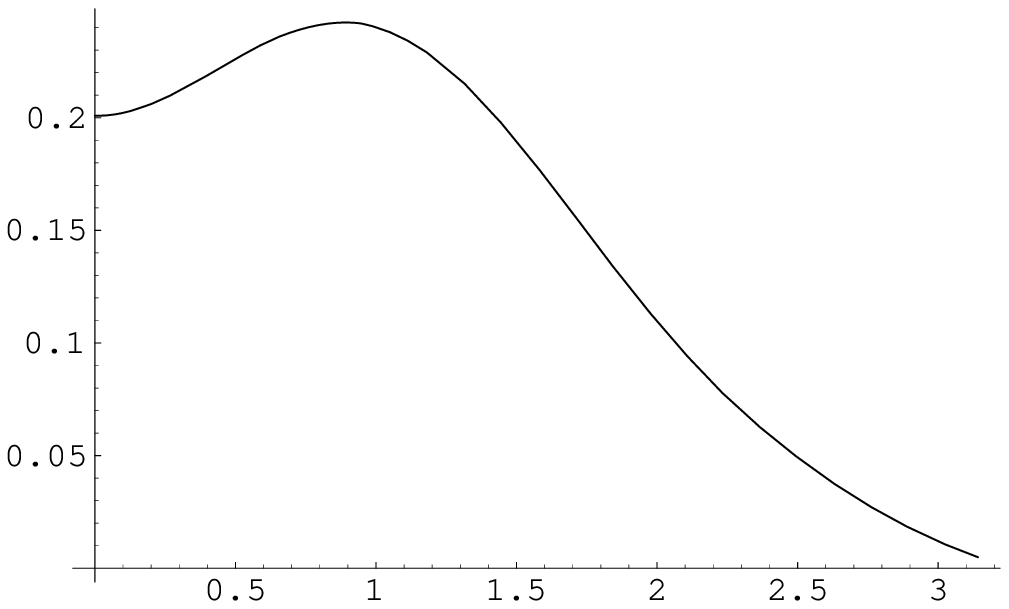}
\hspace{0.0cm} $\Theta \rightarrow$
  \rotatebox{90}{\hspace{1.0cm} $h_m \rightarrow$}
\includegraphics[height=.15\textheight]{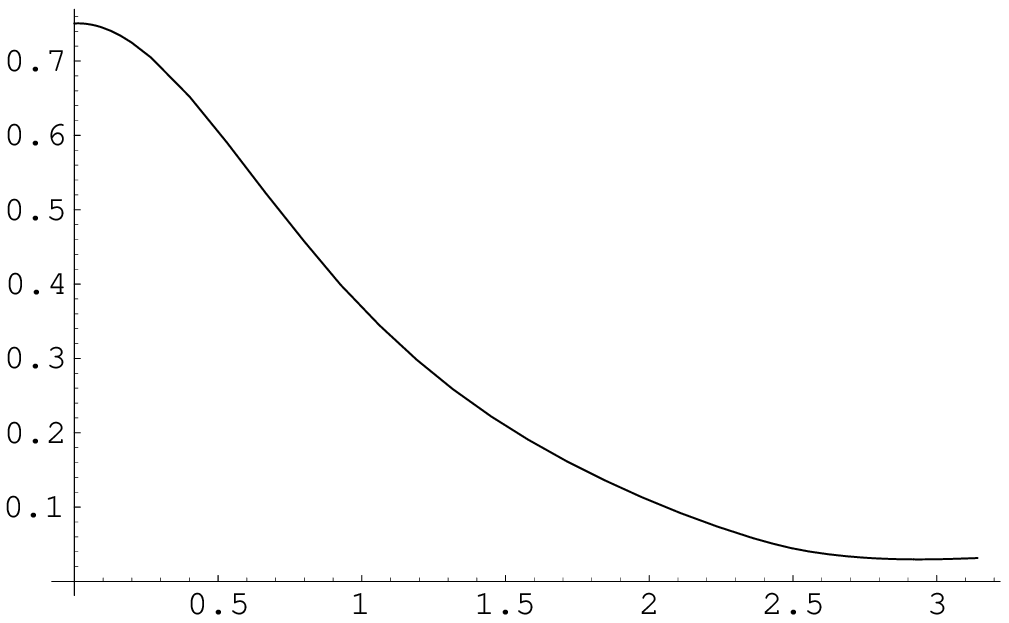}
\hspace{0.0cm} $\Theta \rightarrow$\\
\rotatebox{90}{\hspace{1.0cm} $h_m \rightarrow$}
\includegraphics[height=.15\textheight]{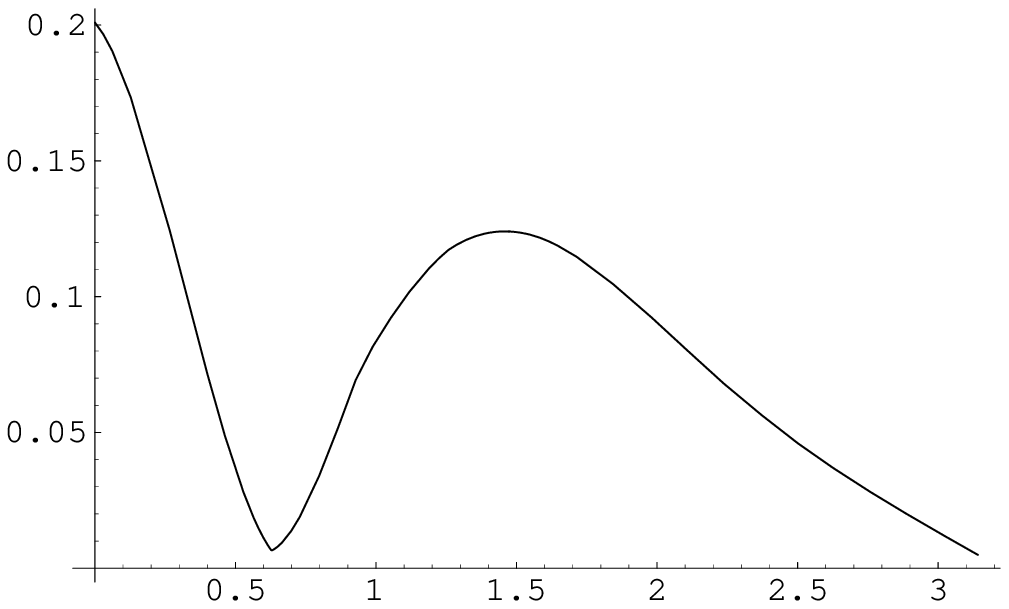}
\hspace{0.0cm} $\Theta \rightarrow$
 \rotatebox{90}{\hspace{1.0cm} $h_m \rightarrow$}
\includegraphics[height=.15\textheight]{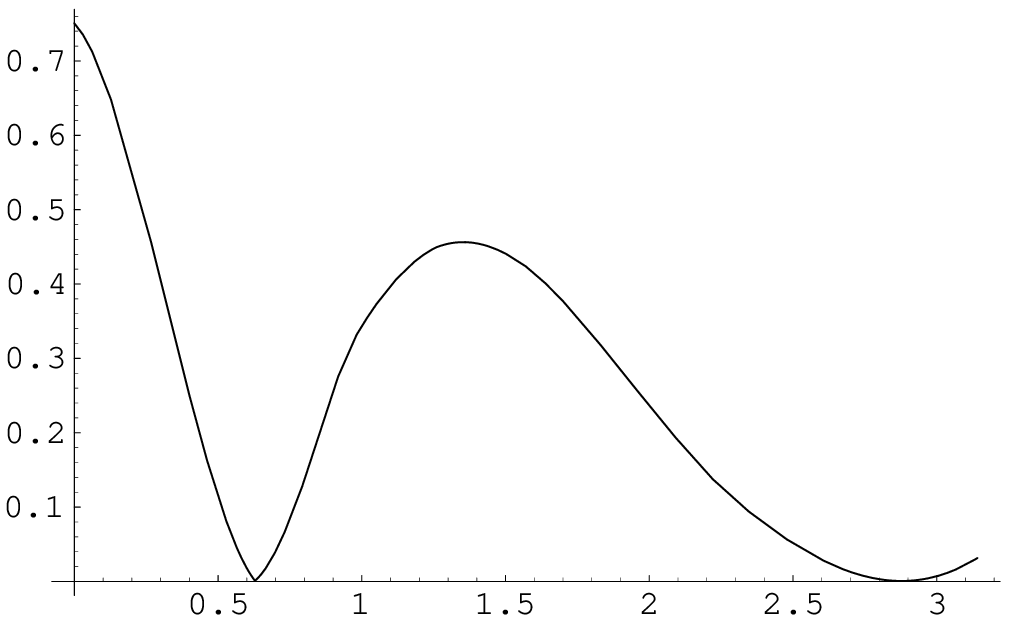}
\hspace{0.0cm} $\Theta \rightarrow$\\
\rotatebox{90}{\hspace{1.0cm} $h_m \rightarrow$}
\includegraphics[height=.15\textheight]{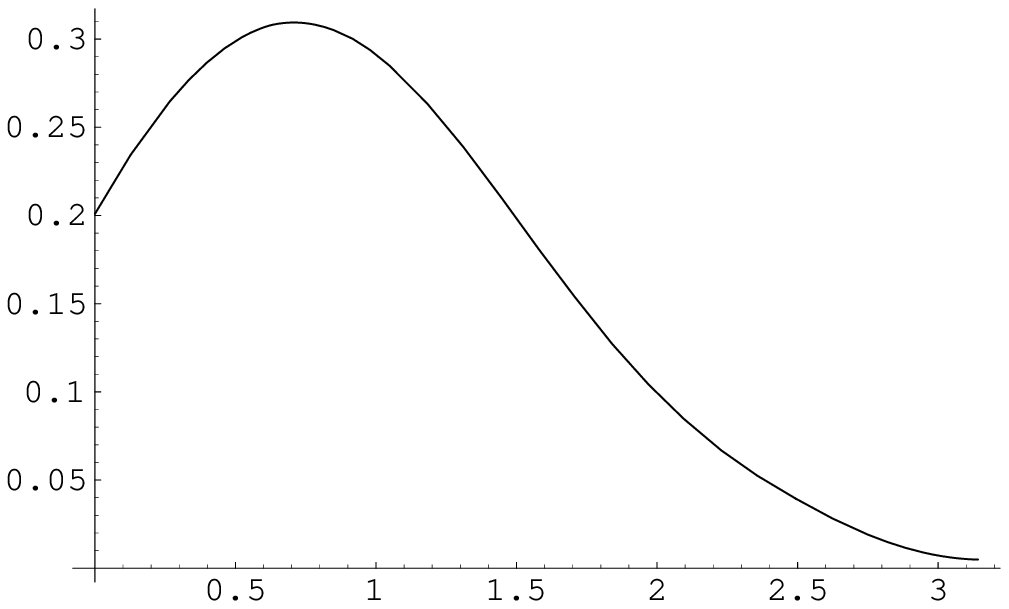}
\hspace{0.0cm} $\Theta \rightarrow$
 \rotatebox{90}{\hspace{1.0cm} $h_m \rightarrow$}
\includegraphics[height=.15\textheight]{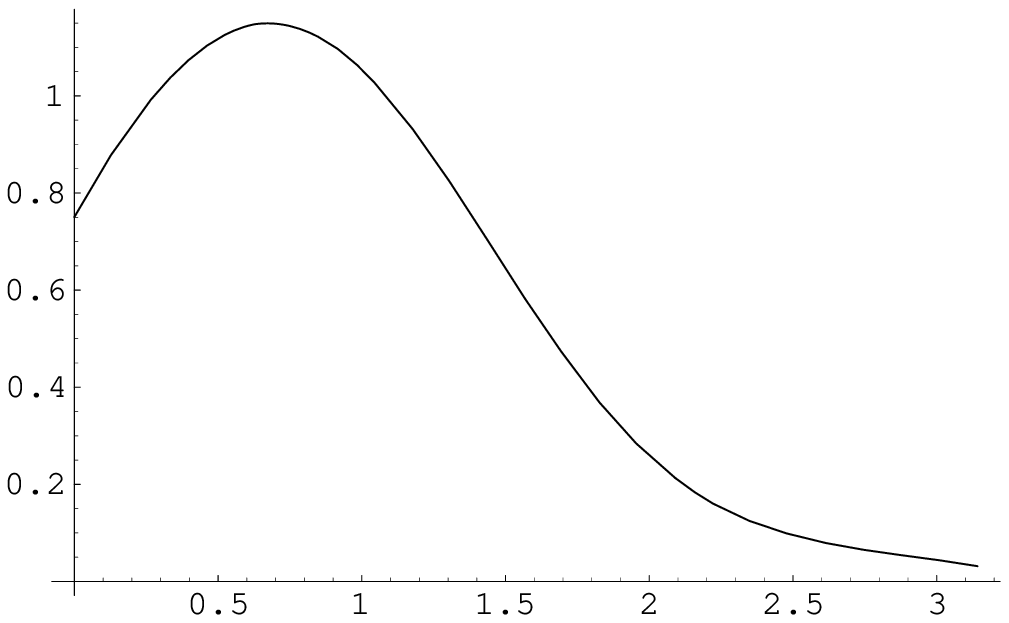}
\hspace{0.0cm} $\Theta \rightarrow$
 \caption{ The same as in Fig.
\ref{hdir19} for A=127.
 \label{hdir127}
 }
\end{center}
\end{figure}

We sometimes prefer to use the parameters
$\kappa=\frac{t_{dir}}{t}$ and $h_m$, since, being ratios, are
expected to be
 less dependent on the parameters of the theory. We first exhibit the
dependence of $h_m$ on the angle $\Theta$ for an LSP mass of $m_{\chi}=100GeV$
in Figs \ref{hdir19} and \ref{hdir127}. Then we exhibit the dependence of the
parameters $t$, $h$, $\kappa,h_m$, and $\alpha_m$,
which are essentially independent of
the LSP mass  for target $A=19$, in
 Table \ref{table1.gaus} (for the other light systems the results are
almost identical).

\begin{figure}
\begin{center}
 \rotatebox{90}{\hspace{1.0cm} $\kappa \rightarrow$}
\includegraphics[height=.15\textheight]{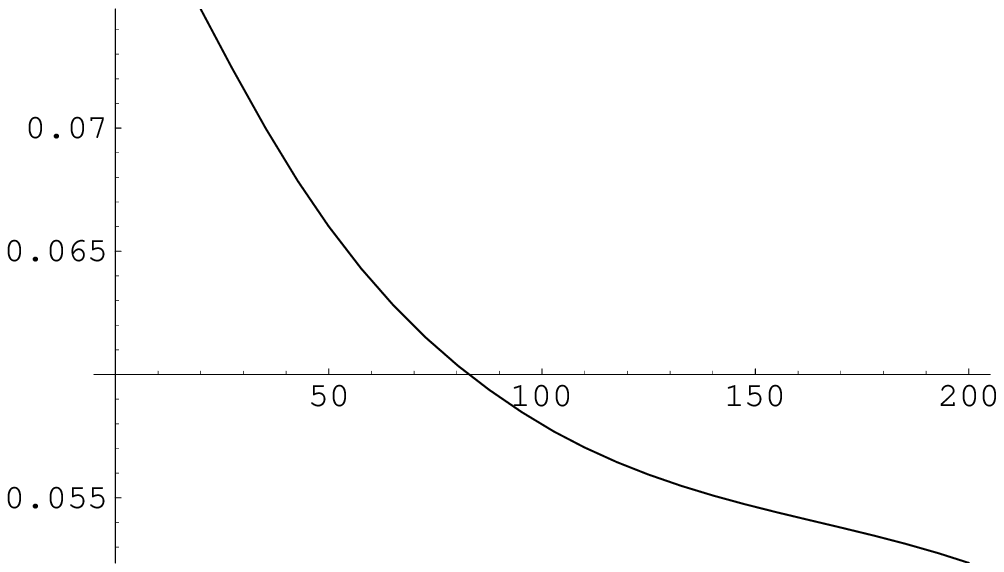}
\hspace{0.0cm} $m_{\chi} \rightarrow$ GeV
 \rotatebox{90}{\hspace{1.0cm} $\kappa \rightarrow$}
\includegraphics[height=.15\textheight]{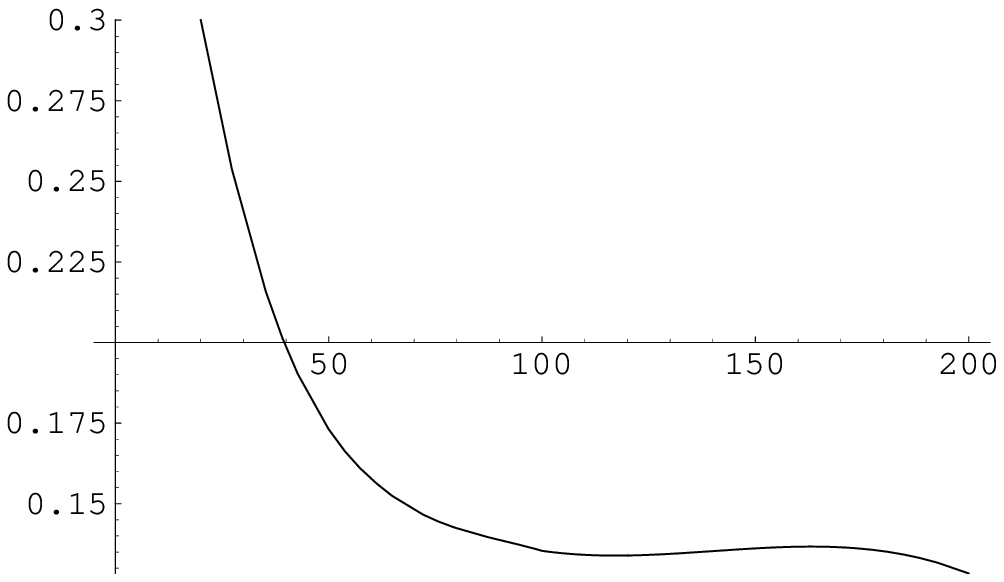}
\hspace{0.0cm} $m_{\chi} \rightarrow$ GeV
 \caption{ The parameter
$\kappa$ as a function of the LSP mass in the case of the $A=127$
system, associated with the Maxwellian velocity distribution
($\lambda=0$) and $Q_{min}=0$. On the left we show the results for
$\Theta=\pi/2$ and on the right $\Theta=\pi$ (the results are
independent of $\Phi$).  The rate for $\Theta=0$ is negligible.
 \label{k.127}
}
\end{center}
\end{figure}

\begin{figure}
\begin{center}
\rotatebox{90}{\hspace{1.0cm} $h_m \rightarrow$}
\includegraphics[height=.15\textheight]{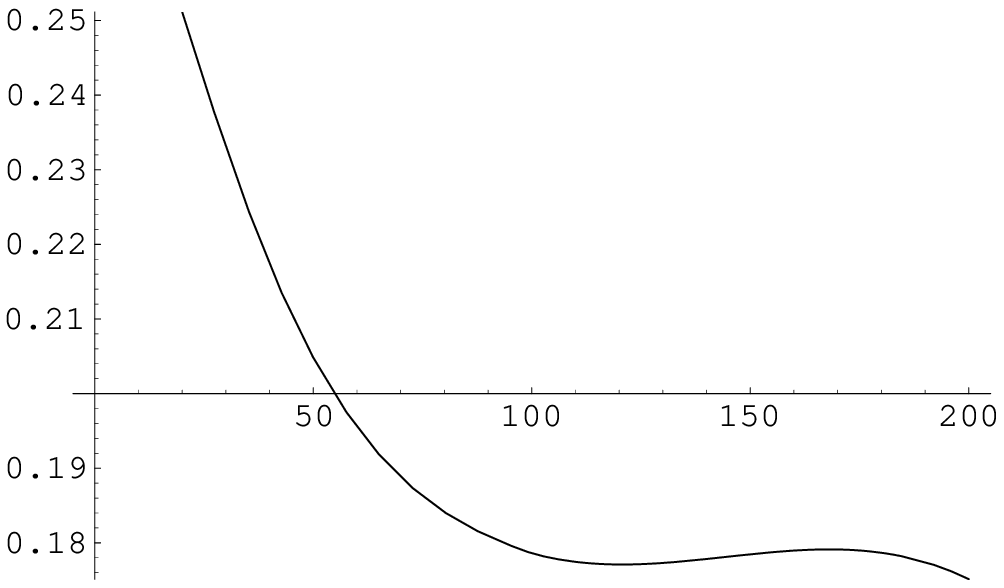}
\hspace{0.0cm} $m_{\chi} \rightarrow$ GeV
\rotatebox{90}{\hspace{1.0cm} $h_m \rightarrow$}
\includegraphics[height=.15\textheight]{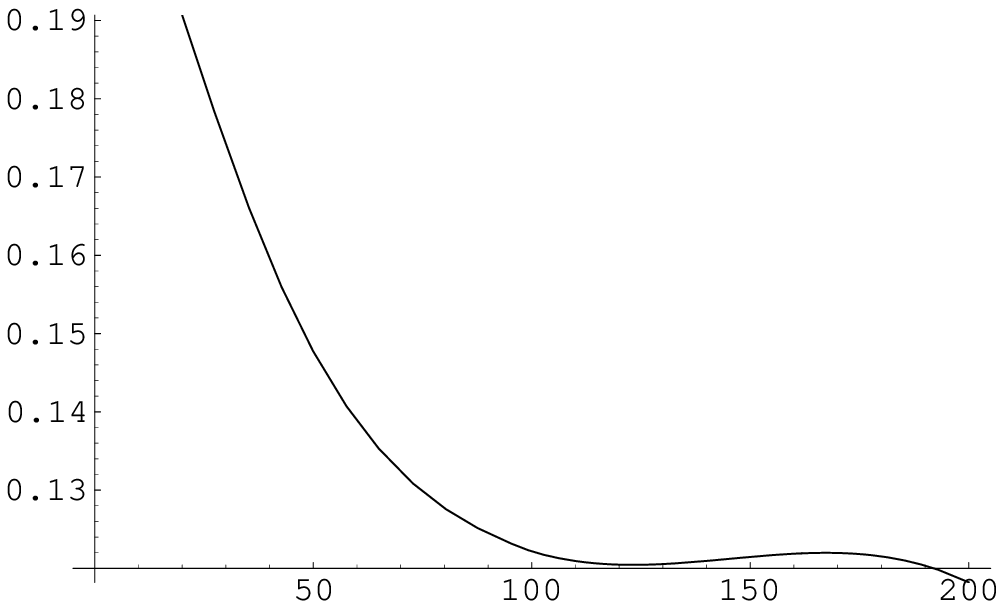}
\hspace{0.0cm} $m_{\chi} \rightarrow$ GeV \\
\rotatebox{90}{\hspace{1.0cm} $h_m \rightarrow$}
\includegraphics[height=.15\textheight]{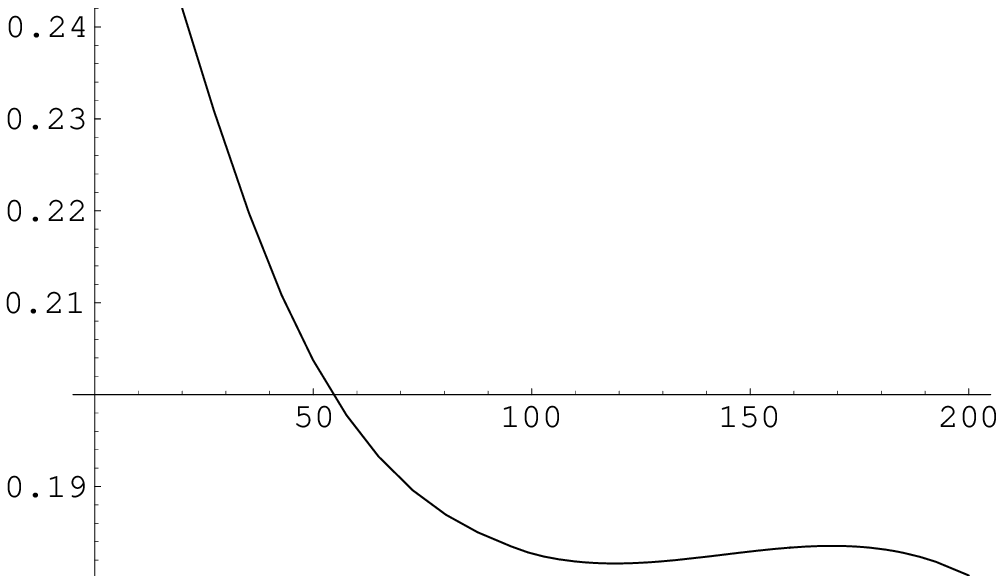}
\hspace{0.0cm} $m_{\chi} \rightarrow$ GeV
\rotatebox{90}{\hspace{1.0cm} $h_m \rightarrow$}
\includegraphics[height=.15\textheight]{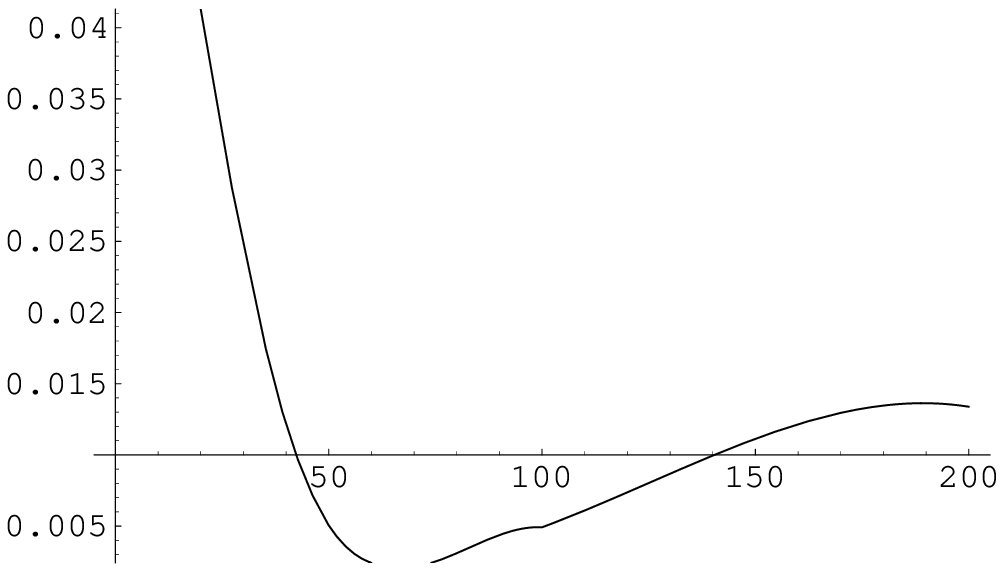}
\hspace{0.0cm} $m_{\chi} \rightarrow$ GeV
  \caption{
The modulation amplitude $h_m$.
 From left to right and top to bottom the graphs correspond
to $\pm x,+y,-y,-z$.
 Otherwise the notation is the same as in Fig \ref{k.127}.
 \label{h.127}
 }
\end{center}
\end{figure}

The asymmetry is quite large. For a Gaussian velocity distribution we find:
$$As=\frac{R(-z)-R(+z)}{R(-z)+R(+z)}\approx 0.97$$ In the other directions
it depends on the phase of the Earth and is equal to almost twice
 the modulation.
For a heavier nucleus the situation is a bit complicated. Now the
parameters $\kappa$ and $h_m$ depend on the LSP mass. The situation is
 exhibited in Figs \ref{k.127} and \ref{h.127}. The asymmetry and the shift
in the phase of the Earth are similar to those of the $A=19$ system.
\\In the results shown we only considered the rotational velocity
of the sun around the center of the galaxy. We do not, however,
expect a significant modification, if the other components of the
sun`s velocity, which are an order of magnitude smaller, are
included.
\section{Conclusions}
 It is well known that, in the
coherent case, only in a small segment of the allowed
 parameter space the rates are above
 the present experimental goals ~\cite{Gomez,ref2,ARNDU}, which of course
may  be improved  by two or three orders of magnitude in the
planned experiments \cite{CDMS}-\cite{GENIUS}. In the case of the
spin contribution only in models with large higgsino components of
the LSP one can obtain rates, which may be presently detectable,
but in this case, except in special models, the bound on the relic
LSP abundance may be difficult to respect. Anyway it appears that
in both cases the expected rates
 are small. It may, therefore, be necessary to exploit any
characteristic experimental signatures, which may reduce the
formidable backgrounds at such low counting rates.

 In the present
paper we considered the spin induced rates and examined the
following signatures:
\begin{itemize}
\item  Correlation of the event rates with the motion of the Earth
(modulation effect) \item Angular correlation of the directional
rates with the direction of motion  of the sun as well as their
seasonal variation. \\
Such experiments are currently under way, like
the UKDMC DRIFT PROJECT experiment \cite{UKDMC}, the Micro-TPC Detector of the
Kyoto-Tokyo collaboration \cite {KYOTOK} and the TOKYO experiment
\cite {TOKYO}.
\end{itemize}
Let us first consider the conventional experiments and focus on
the relative parameters $t=R/\bar{R}$ and the modulation (seasonal
variation) amplitude $h$, normalized to zero when the motion of
the Earth is ignored. In the case of light nuclear targets they
are essentially independent of the LSP mass, but they depend on
the energy cutoff, $Q_{min}$. For $Q_{min}=0$ they are exhibited
in Table \ref{table1.gaus}. They are essentially the same for both
the coherent and the spin modes.
 For intermediate and heavy nuclei they depend on the
LSP mass \cite{JDV03}.
 It is clear that among the light targets, from the point of view of the static spin
 matrix elements, the
most favored system \cite{DIVA00} is the $A=19$, since the spin
matrix elements are both large and very reliably calculated. We
should keep in mind, however,  that  for heavy LSP the reduced
mass can be large in the case of a heavy nucleus like $^{207}Pb$.
 The increase of the rates caused by the increase of the reduced mass
 may very well compensate for the smallness of the spin matrix elements. As
a matter of fact in the case of $^{127}I$ we see that, for an LSP
mass greater than $100~GeV$, the obtained rate is larger than that
of $^{19}F$ in spite of the fact that the
 dominant isovector static spin matrix elements are in the opposite order.
 We should not forget, though, that the calculated spin matrix elements in
the case of the $A=19$ system are more reliable.

In the case of the directional experiments, i.e  when only nuclei
recoiling in a certain direction are counted,  we can summarize
our results as follows:
\begin{itemize}
\item The angular dependence of the  rate.\\
The ratio of the directional rate divided by that of the usual
rate , given by $\approx \kappa/(2\pi)$, is smaller than $1/5\pi$.
Such a big loss in the rate may very well be compensated by the
important experimental signature associated with the fact that the
factors
 $\kappa$ depend on the direction of observation. In the case
 of a Gaussian velocity
distribution this factor is the largest, $\kappa\approx 0.4$,
 when the nucleus is recoiling
opposite to the direction of motion of the sun.
 $\kappa$ is the smallest in the direction of
the sun's motion (least favored direction) and in between for the
other directions. In other words the directional rate is strongly
peaked in the direction opposite to the velocity of the sun and
the resulting event asymmetry between these two directions is very
large. In a plane perpendicular to the sun's direction of motion
$\kappa$ takes intermediate values, $\kappa \approx 0.08$. The
asymmetry between one direction and its opposite is in this case
almost zero, if the motion of the Earth is neglected. Anyway the
study of such angular correlations may play a  role in confirming
any observed neutralino events, since there may appear seasonal
effects, which can mimic the small modulation in the conventional
experiments.

 \item Seasonal variation (modulation) of the directional rates.\\
We have shown that the directional rates can, in addition, exhibit
seasonal variation (modulation). In the most favored direction the
modulation is not very large, but still it is
 three times larger compared to that expected in the standard non directional
 experiments. This gain is , perhaps, not
 big enough to compensate for the reduction in the number of the events.
 In a plane perpendicular to the sun's direction of motion,
 however, the modulation
is quite large (see Table \ref{table1.gaus}).
 The experimentalists will themselves decide what use, if any, to make
of these predictions. A simple argument indicates that they may be
useful. Taking, as an example, events in the
 x-direction (radial galactic direction), we can see that the modulation
 signal will be reduced by
 a factor of $(2\pi/0.08)(0.272/0.02)=5.8$
 The background will be reduced by a factor of $\sqrt{(2\pi/0.08)}=8.9$. Thus one
 expects a gain of $1.5$.\\  Furthermore the modulation in this plane
  is  characterized by a very
 interesting seasonal pattern, depending on the angle of observation
 (see Table \ref{table1.gaus}). Perhaps this also can be exploited by the
 experimentalists.
\end{itemize}
 The predicted reductions in the rates of the directional experiments
 compared to the standard experiments, at a first sight, may be
  seen as an obstacle to be overcome for the benefit
 of the above good signatures. This may be true  in
  the case of some of the planned experiments
\cite{TOKYO}, which intend to use organic detectors capable of
making observations in only one predetermined direction  for each
run. Quite clearly, however, the directional observations are
going to offer only advantages in the case of experiments using
TPC detectors
 \cite{UKDMC,KYOTOK}.
 The TPC counters can simultaneously  register all events.
  If something interesting is
found, the analysis can be made directionally to reject possible
background events.

We finally hope that, in spite of the  reduction in the predicted
rates, the signatures of the directional experiments (large
asymmetry and modulation amplitude) can be exploited by the
experimentalists.

\par
Acknowledgments: This work was supported in part by the European
Union under the contracts RTN No HPRN-CT-2000-00148 and TMR No.
ERBFMRX--CT96--0090. Part of this work was performed in LANL. The
author is indebted to Dr Dan Strottman for his support and
hospitality and to Hiro Ejiri for his useful comments on the
preparation of the manuscript. He is also grateful to Y.
Giomataris and K. Zioutas for bringing to his attention the great
opportunities offered by the TPC detectors.

\begin{table}[t]
\caption{ A summary of the parameters $C^n_i$ of the model of
Chattopadhyay and Roy and the relevant renormalization factors. }
\label{table.Roy}
\begin{center}
\begin{tabular}{|l|ccc|r|}
\hline \hline
& & & & \\
n/i &  3&   2  & 1&comment\\
\hline \hline
1 &1& 1& 1& SUGRA (Bino)\\
24 &1& -3/2& -1/2& Bino\\
75&1& 3& -5& Zino\\
100&1& 2& 10& Higgsino\\
\hline \hline
&25/9&25/30&25/60&Ren. factor\\
\hline \hline
\end{tabular}
\end{center}
\end{table}
\begin{table}[t]
\caption{
The static spin matrix elements for various nuclei.
For light nuclei the calculations are from Divari et al (see text) . For $^{127}I$ the
results are from Ressel and Dean (see text) (*)  and  the
 Jyvaskyla-Ioannina collaboration (private communication)(**).
 For $^{207}Pb$ they were obtained previously (see text).
}
\label{table.spin}
\begin{center}
\begin{tabular}{lrrrrrr}
 &   &  &  &  &   & \\
 & $^{19}$F & $^{29}$Si & $^{23}$Na  & $^{127}I^*$ & $ ^{127}I^{**}$ & $^{207}Pb^+$\\
\hline
    &   &  &  &  &    \\
$[\Omega_{0}(0)]^2$         & 2.610   & 0.207  & 0.477  & 3.293   &1.488 & 0.305\\
$[\Omega_{1}(0)]^2$         & 2.807   & 0.219  & 0.346  & 1.220   &1.513 & 0.231\\
$\Omega_{0}(0)\Omega_{1}(0)$& 2.707   &-0.213  & 0.406  &2.008    &1.501&-0.266\\
$\mu_{th} $& 2.91   &-0.50  & 2.22  &    &\\
$\mu_{exp}$& 2.62   &-0.56  & 2.22  &    &\\
$\frac{\mu_{th}(spin)}{ \mu_{exp}}$& 0.91   &0.99  & 0.57  &    &  &\\
\end{tabular}
\end{center}
\end{table}
\begin{table}[t]
\caption{ The parameters $t$, $h$, $\kappa,h_m$ and $\alpha_m$ for the
 isotropic Gaussian
 velocity distribution and $Q_{min}=0$. The results presented are  associated
 with the spin contribution, but those
 for the coherent mode are similar. The results shown are for the light
systems. For intermediate and heavy nuclei there is a dependence on the LSP mass. $+x$ is
 radially  out of the galaxy ($\Theta=\pi/2,\Phi=0$), $+z$ is in the sun's
 direction of motion ($\Theta=0$) and
$+y$ is vertical to the plane of the galaxy ($\Theta=\pi/2,\Phi=\pi/2$) so that
 $(x,y,z)$ is right-handed. $\alpha_m=0,1/2,1,3/2$ means
that the maximum occurs on the 2nd of June, September, December and March
 respectively.
\label{table1.gaus}}
\begin{center}
\begin{tabular}{lrrrrrr}
& & & & & &      \\
type&t&h&dir &$\kappa$ &$h_m$ &$\alpha_m$ \\
\hline
& & & & & &      \\
& &&+z        &0.0068& 0.227& 1\\
dir& & &+(-)x      &0.080& 0.272& 3/2(1)\\
& & &+(-)y        &0.080& 0.210& 0 (1)\\
& & &-z         &0.395& 0.060& 0\\
\hline
all&1.00& & && & \\
all& & 0.02& & & & \\
\hline
\end{tabular}
\end{center}
\end{table}
\end{document}